\documentclass[fleqn,usenatbib]{mnras}

\usepackage{newtxtext,newtxmath}

\usepackage[T1]{fontenc}

\DeclareRobustCommand{\VAN}[3]{#2}
\let\VANthebibliography\thebibliography
\def\thebibliography{\DeclareRobustCommand{\VAN}[3]{##3}\VANthebibliography}

\usepackage{subcaption}
\usepackage{graphicx}	
\usepackage{amsmath}	

\DeclareMathOperator{\tr}{tr}

\DeclareMathOperator\diag{diag}

\title[Rescaling Transforms for Spherical Flows]{Rescaling Transforms for Local Models of Spherical Flows}

\author[Elliot M. Lynch and Guillaume Laibe]{
Elliot M. Lynch,$^{1}$\thanks{E-mail: elliot.lynch@ens-lyon.fr}
Guillaume Laibe,$^{1}$
\\
$^{1}$ENS de Lyon, CRAL UMR5564, Universite Claude Bernard Lyon 1, CNRS, Lyon 69007, France
}

\date{Accepted XXX. Received YYY; in original form ZZZ}

\pubyear{2025}

\begin{document}
\label{firstpage}
\pagerange{\pageref{firstpage}--\pageref{lastpage}}
\maketitle

\begin{abstract}
 Previously we developed a local model for a spherically contracting/expanding gas cloud that can be used to study turbulence and small scale instabilities in such flows. In this work we generalise the super-comoving variables used in studies of cosmological structure formation to our local spherical flow model, which make it significantly easier to derive analytical solutions and analyse the interactions of more complex flows with the background. We show that a wide class of solutions to the local spherical flow model can be obtained via a mapping from the corresponding solutions in regular Cartesian flows. The rescaling of time in the transformation results in a modification of the linear instabilities that can occur in spherical flows, causing them to have a time dependent growth rate in the physical time coordinate, and can prevent slower instabilities from operating. Finally, we show that the small scale flows in isotropic contraction/expansion can be mapped directly to Cartesian, inviscid, incompressible hydrodynamics, meaning that one expects a form of rescaled Kolmogorov-turbulence at the small scale of isotropically contracting/expanding flows. 
\end{abstract}




\section{Introduction}

Spherically contracting/expanding flows are a classic problem in astrophysics, with diverse applications from star-formation \citep{Larson69,Penston69,Shu77,Hunter77,Foster93} to supernova explosions \citep{Sedov46,Taylor46,Taylor50,Bethe58}. Isotropic contracting/expanding flows are of interest in cosmological structure formation \citep{Shandarin80,Martel98,Voit96} to turbulence in the interstellar medium \citep{Robertson12,Hennebelle21} and giant molecular clouds \citep{Robertson12}. Up to supercluster scales such isotropic contracting/expanding flows can be dealt with within the context of Newtonian-dynamics \citep{Shandarin80,Robertson12}. 

 Both the local dynamics of the spherically contracting/expanding flow and the (Newtonian) isotropic contraction/expansion can be studied within the framework of local contracting/expanding boxes, which consist of a periodic domain where the background collapse is absorbed into the time-dependent, but spatially homogeneous, box geometry. Local models for a uniformly expanding magnetohydrodynamical (MHD) flow were first developed to study the solar wind \citep{Velli92,Grappin93,Grappin96} and have been generalised to an accelerated expansion by \citet{Tenerani17}. In the astrophysical hydrocode \textsc{athena++} \citep{Stone20}, a numerical implementation for an isotropically contracting hydrodynamical flow was developed by \citet{Robertson12}, and for an anisotropic MHD flow by \citet{Sun23}. \citet{Lynch23} rederived the local model for a spherical hydrodynamical flow, with particular attention to ensuring that the pressure gradient terms are compatible with periodic boundary conditions, along with determining the conditions for the eventual breakdown of the local model expected in contracting flows. \citet{Lynch23} also derived several linear and nonlinear solutions that are useful for understanding the fundamental physics of such models and for testing numerical implementations. \citet{Xu24} developed a numerical implementation of the model of \citet{Lynch23} and showed that their numerical implementation recovers these properties and analytical solutions.

In this work we present a complementary approach to modelling such local spherical flows. Here, instead of making use of a numerical implementation of the local model we instead show that a wide class of solutions to the local spherical flow model can be obtained via a mapping from the corresponding solutions in regular Cartesian flows. A similar mapping has been found between adiabatic cosmological (i.e. isotropic) flows for a monatomic gas and the standard Cartesian box, referred to as super-comoving coordinates \citep{Shandarin80,Martel98} that have been implemented in astrophysical hydrocodes \citep[e.g. \textsc{ramses}][]{Teyssier02}. This means the adiabatic version of the simulations of \citet{Robertson12} can be obtained from a standard decaying turbulence simulations \citep[e.g.][]{Sytine00}, suitably rescaled. In the cosmological context these super-comoving variables have been used to generalise blast wave solutions to isotropically expanding flows \citep{Voit96}.

One property of this class of rescaling, which has important physical consequences, is the necessity of rescaling time. This means many linear instabilities, when rescaled for an expanding flow, cannot operate as they require an infinite physical time to undergo a single e-folding. Other instabilities in contracting flows may not have time to operate before the flow becomes rotationally supported. The instabilities that can operate are exponentially growing in the rescaled time coordinate, meaning they have a time-varying growth rate when mapped back to physical time.

The plan of the paper is as follows. In Section \ref{rescaling derivation} we derive the box rescaling and key properties common to all of the rescalings. Section \ref{particular rescalings} details specific rescalings that simplify some of the local model properties. Section \ref{incomp limit} demonstrates that spherical flows posses an incompressible limit, important for understanding the flow on small scales. In Section \ref{Rescaled time} we discuss the effects of the rescaled time and the termination of the solution after finite rescaled time, either as a result of a breakdown of the conditions for the local model to be valid, or due to requiring an infinite physical time to have passed upon the inversion of the rescaling. These consideration are particularly important when discussing instabilities in the rescaled flow. Section \ref{2d sols} demonstrates the utility of the rescaling by showing how solutions and instabilities from 2D hydrodynamics can be rescaled to the spherical flow problem. Section \ref{discuss} discusses how the rescaling could be used to aid in the study of turbulence in such local models in the future and we present our conclusions in Section \ref{conc}. Finally Appendix \ref{common flows} gives the rescalings for several example background flows, while Appendices \ref{apprent vort equation}, \ref{d dimensional conformal} and \ref{linear incomp} contain supplementary mathematical derivations.

\section{Deriving the Rescaling Transform for the Local Spherical Flow Equations} \label{rescaling derivation}

\subsection{Deriving the rescaled equations}

For an ideal fluid the local spherical flow equations can be written in the form \citep{Lynch23}

\begin{align}
 D \rho &= - \rho \left[\Delta + \boldsymbol{\partial} \cdot \mathbf{v} \right] \quad , \\
 \rho D \mathbf{v} &+ \rho \mathbf{g}^{-1} \dot{\mathbf{g}} \mathbf{v} = - \mathbf{g}^{-1} \boldsymbol{\partial} p \quad , \\
 D p &= - \gamma p \left[\Delta + \boldsymbol{\partial} \cdot \mathbf{v} \right] \quad ,
\end{align}
where $\rho$ is the fluid density; $\mathbf{v}$ is the fluid velocity relative to the background flow; $p$ is the fluid pressure; $\gamma$ the ratio of specific heats; $\boldsymbol{\partial}$ denotes a spatial partial derivative; $D = \partial_t + \mathbf{v} \cdot \boldsymbol{\partial}$ is the Lagrangian time derivative with respect to the relative flow; $\Delta$ is the velocity divergence of the background flow and $\mathbf{g}$ is the metric tensor of the local model, represented as a diagonal matrix of the respective squared lengths of the three directions

\begin{equation} 
 \mathbf{g} = \begin{pmatrix}
 L_1^2 (t) & 0 & 0 \\
 0 & L_2^2 (t) & 0 \\
 0 & 0   & L_3^2 (t) 
\end{pmatrix} \quad ,
\end{equation}
where we have kept the three lengthscale independent, similar to \citet{Sun23}, to be applicable to a wider range of problems. The determinant of $\mathbf{g}$, $|\mathbf{g}|$ is related to the $3$ lengthscales by $|\mathbf{g}|^{1/2} = L_1 L_2 L_3$ and the velocity divergence in the background flow is related to this determinant by

\begin{equation} 
 \Delta = |\mathbf{g}|^{-1/2} \partial_t |\mathbf{g}|^{1/2} = \frac{\dot{L}_1}{L_1} + \frac{\dot{L}_2}{L_2}  + \frac{\dot{L}_3}{L_3} .
\end{equation}

Due to explicitly solving the thermal energy equation, the vorticity equation is generalised from that presented in \citep{Lynch23} to include a baroclinic term:

\begin{equation}
 D \boldsymbol{\omega} + \boldsymbol{\omega} [\Delta + \boldsymbol{\partial} \cdot \mathbf{v}] - \boldsymbol{\omega}  \cdot \boldsymbol{\partial} \mathbf{v} = \rho^{-2} |\mathbf{g}|^{-1/2} (\boldsymbol{\partial} \rho \times \boldsymbol{\partial} p) \quad ,
\end{equation}
With the vorticity related to the velocity and metric tensor by $\boldsymbol{\omega} = |\mathbf{g}|^{-1/2} (\boldsymbol{\partial} \times \mathbf{g} \mathbf{v})$. 

\begin{figure*}
\includegraphics[trim=0 360 0 10, clip, width=\linewidth]{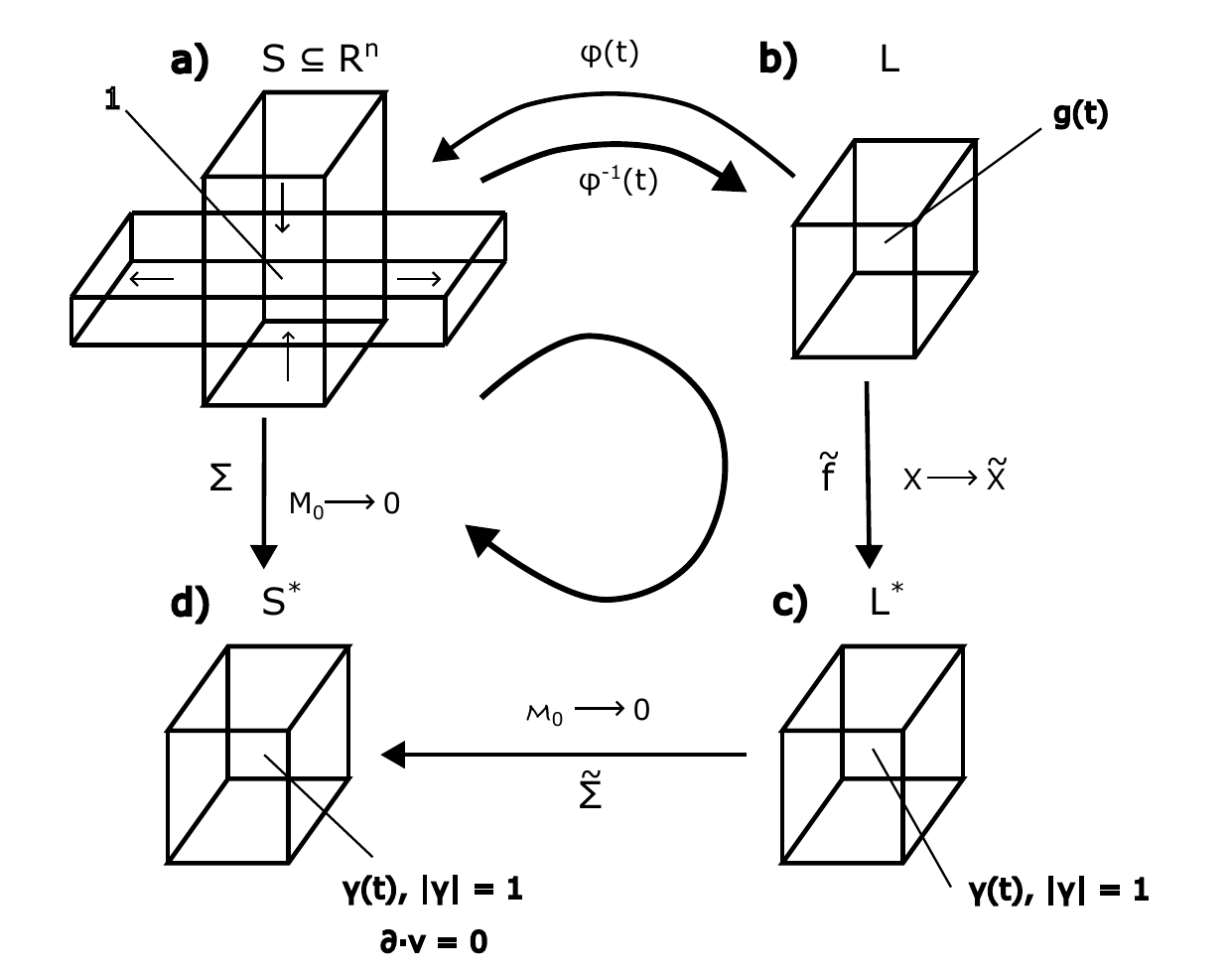}
\caption{Diagram showing the different transforms/spaces used this paper. a) Shows the domain, $S \subseteq \mathbb{R}^3$, in the global frame, where the size and shape of the domain changes as a result of the background flow. b) Shows the local model, $L$, with metric $\mathbf{g}$ and time-dependent coordinate chart $\varphi$ that maps the local model to $S$. c) Shows one of the rescaled local models, $L^{*}$, obtained by introducing rescaled variables analogous to the super-comoving coordinates of \citet{Shandarin80,Martel98}, with new metric $\boldsymbol{\gamma}$ that has determinant $1$. d) Shows the incompressible limit of c), $S^{*}$,obtained by taking the rescaled Mach number, $\mathcal{M}_0$, to zero. By composition of maps, $\Sigma = \tilde{\Sigma} \circ \tilde{f} \circ \varphi^{-1}$, we thus have a way of obtaining the incompressible limit of the original problem, a), despite the presence of compression in the background flow.}
\label{mapping cartoon}
\end{figure*}

Introducing a generic time-dependent rescaling of the variables of the form

\begin{equation}
 \rho = \rho_0 (t) \tilde{\rho} , \quad \mathbf{v} = v_0 (t) \tilde{\mathbf{v}} , \quad p = p_0 (t) \tilde{p} ,
\end{equation}
where $\rho_0$, $v_0$ and $p_0$ are functions of time only, to be specified, along with a rescaled time coordinate of the form

\begin{equation}
 \tau = \int v_0 \, d t.
\end{equation} 
The Lagrangian time derivative rescales as

\begin{equation}
 D = \partial_t + \mathbf{v} \cdot \boldsymbol{\partial} = v_0 (\partial_{\tau} + \tilde{\mathbf{v}} \cdot \boldsymbol{\partial}) = v_0 \tilde{D} ,
\end{equation}
while the spatial derivatives remain unchanged. Setting $\rho_0 = |\mathbf{g}|^{-1/2}$, which absorbs the background velocity divergence in the continuity equation,  then, after some rearranging, the fluid equations become

\begin{align}
 \tilde{D} \tilde{\rho} &= - \tilde{\rho} \, \boldsymbol{\partial} \cdot \tilde{\mathbf{v}} , \label{rescale rho eq} \\
 \tilde{\rho} \tilde{D} \tilde{\mathbf{v}} + \tilde{\rho} v_0^{-1} \left(\frac{\dot{v}_0}{v_0} \mathbf{1} + \mathbf{g}^{-1} \dot{\mathbf{g}} \right) \tilde{\mathbf{v}} &= - \frac{p_0 |\mathbf{g}|^{1/2} }{v_0^2} \mathbf{g}^{-1} \boldsymbol{\partial} \tilde{p} \\
\tilde{D} \tilde{p} + \gamma \tilde{p} \, \boldsymbol{\partial} \cdot \tilde{\mathbf{v}} &= \Gamma, \label{rescale p eq}
\end{align}
where we have introduced an effective heating/cooling rate

\begin{equation}
 \Gamma = \tilde{p}/t_{\rm heat} = -v_0^{-1} \left( \frac{\dot{p}_0}{p_0} + \gamma \Delta \right) \tilde{p} ,
\end{equation}
with $|t_{\rm heat}|$ the characteristic heating/cooling timescale in the rescaled time coordinate. Depending on the problem in question it is possible to choose $v_0$ and $p_0$ such that the rescaled equations (Equations \ref{rescale rho eq}-\ref{rescale p eq}) are simplified dramatically. Figure \ref{mapping cartoon} shows the relationship between the different spaces for one such rescaling (conformal rescaling), introduced in Section \ref{conf rescl}, which absorbs the volumetric part of the background flow into this effective heating term.

\subsection{Physical properties of the rescaled equations} \label{physical properties  rescaled}

The presence of a heating term results in entropy production in the rescaled system, with the entropy, $\tilde{s}$, of the rescaled system evolving according to

\begin{equation}
 \tilde{D} \tilde{s} = \frac{R_{\rm gas}}{(\gamma - 1) t_{\rm heat}} \quad , \label{entropy eq}
\end{equation}
where $R_{\rm gas}$ is the specific gas constant. From this we see that the heating term is not a source of baroclinicity, as a result of having the same spatial variation as the pressure.
 
As noted in \citet{Lynch23}, $\tilde{\boldsymbol{\omega}} = |\mathbf{g}|^{1/2} \boldsymbol{\omega}$ is conserved by the local flow equations. In terms of the rescaled variables, and including baroclinic terms, the rescaled vorticity equation becomes

\begin{equation}
 \tilde{D} \tilde{\boldsymbol{\omega}} - \tilde{\boldsymbol{\omega}} \cdot \boldsymbol{\partial} \tilde{\mathbf{v}} + \tilde{\boldsymbol{\omega}} \boldsymbol{\partial} \cdot \tilde{\mathbf{v}} = \frac{p_0 |\mathbf{g}|^{1/2}}{v_0} \tilde{\rho}^{-2} (\boldsymbol{\partial} \tilde{\rho} \times \boldsymbol{\partial} \tilde{p}) , \label{vort ew}
\end{equation}
where the rescaled vorticity is related to the rescaled velocity by $\tilde{\boldsymbol{\omega}} = v_0 (\boldsymbol{\partial} \times \mathbf{g} \tilde{\mathbf{v}})$. Another important related quantity is the ``apparent'' vorticity $\boldsymbol{\omega}_{\mathbf{1}} = \boldsymbol{\partial} \times \tilde{\mathbf{v}}$, i.e. the vorticity one would compute from the rescaled quantities if one were ignorant of the geometric factors and is the vorticity of the analogue problem in the Cartesian box. This apparent vorticity is not conserved (even absent baroclinic terms) in the presence of anisotropy in the background flow. Appendix \ref{apprent vort equation} derives the evolutionary equation for this apparent vorticity, and describes its interaction with the background flow.

\subsection{Coupling of effective heating/cooling to compressive motions}

One property of the effective heating/cooling term is that it is primarily coupled to compressive motions within the rescaled system. This means that when it acts as a heating term it will tend to increase the amplitude of compressive motions within the box over solenoidal ones. Likewise when it is acting as effective cooling the motion within the box tends to become more solenoidal. To see this one can write Equations \ref{rescale rho eq}-\ref{rescale p eq} in hyperbolic form
\begin{equation}
 \dot{\mathbf{X}} + \mathbf{A}^i \partial_i \mathbf{X} = \mathbf{S} ,
\end{equation}
where $\mathbf{X} = (\tilde{\rho}, \tilde{\mathbf{v}}, \tilde{p})$ is the state vector, $\mathbf{S} = \left(0,v_0^{-1} \left(\frac{\dot{v}_0}{v_0} \mathbf{1} + \mathbf{g}^{-1} \dot{\mathbf{g}} \right) \tilde{\mathbf{v}},\Gamma\right)$ is the source vector. The eigenvalues and eigenvectors of the matrices $\mathbf{A}^i$ determine the different waves in the system (see \citealt{Xu24} for details). The heating function $\Gamma$ is only a source for the left and right sound waves,

\begin{equation}
 \mathbf{X}_i^{\pm} = \begin{pmatrix}
 \tilde{\rho} \\
 \pm\frac{1}{L_i} \sqrt{\frac{\gamma p_0 \tilde{p} |\mathbf{g}|^{1/2}}{\tilde{\rho}  v_0^2 }} \hat{\mathbf{e}}_i \\
 \gamma \tilde{p} 
\end{pmatrix} ,
\end{equation}
as these are the only waves that include a pressure perturbation. The effect of this effective heating function can be seen in the linear wave calculations of \citet{Lynch23}, where it captures the effects of the time-varying envelope of the momentum associated with the sound wave ($P_{\alpha}$ in their notation). This contrasts with the other two linear waves ($P_{\beta}$ and $P_{\gamma}$), whose momenta are integrals of motion.

That the effective heating term has no direct effect on the solenoidal motions within the box can be seen from the vorticity equation (Equation \ref{vort ew}), which contains no source term due to the effective heating.  The only source term present in the rescaled vorticity equation is that due to baroclinicity and, as shown in Section \ref{physical properties  rescaled}, the effective heating is not a source of baroclinicity in the fluid.

\section{Particular rescalings} \label{particular rescalings}

\subsection{Conformal rescaling} \label{conf rescl}

One can introduce a conformal rescaling with $v_0 =|\mathbf{g}|^{-1/3}$ and $p_0 = |\mathbf{g}|^{-5/6}$, which allows one to remove the isotropic part of the collapse and introduce conformal metric, $\boldsymbol{\gamma}$\footnote{Not to be confused with $\gamma$, the ratio of specific heat.}, related to the original metric by  

\begin{equation}
 \mathbf{g} = |\mathbf{g}|^{1/3} \boldsymbol{\gamma} , 
\end{equation}
with the new conformal metric having determinant $|\boldsymbol{\gamma}| = 1$. For isotropic collapses this simplifies greatly with $\boldsymbol{\gamma} = \mathbf{1}$ and corresponds to the super-comoving variables of \citet{Shandarin80} and \citet{Martel98}. The resulting fluid equations simplify to

\begin{align}
 \tilde{D} \tilde{\rho} &= - \tilde{\rho} \, \boldsymbol{\partial} \cdot \tilde{\mathbf{v}} , \label{conformal cont} \\
 \tilde{\rho} \tilde{D} \tilde{\mathbf{v}} + \tilde{\rho} \boldsymbol{\gamma}^{-1}  \boldsymbol{\gamma}_{\tau}  \tilde{\mathbf{v}} &= - \boldsymbol{\gamma}^{-1} \boldsymbol{\partial} \tilde{p} \\
\tilde{D} \tilde{p} + \gamma \tilde{p} \, \boldsymbol{\partial} \cdot \tilde{\mathbf{v}} &= \Gamma , \label{conformal pre}
\end{align}
with the heating function given by

\begin{equation}
 \Gamma = -\frac{1}{2} \left(\gamma  - \frac{5}{3} \right) \frac{\partial_{\tau} |\mathbf{g}|}{|\mathbf{g}|} \tilde{p} \label{heating func} .
\end{equation}
For $\gamma=5/3$ (i.e. monatomic gas) this vanishes. This means for two background flows, with $\gamma=5/3$, characterised by metric tensors $\mathbf{g}_1$ and $\mathbf{g}_2$ respectively such that the two metric tensors are related by a conformal rescaling, then the fluid flows are equivalent under the conformal rescaling symmetry. Isotropic flows with $\gamma=5/3$ are equivalent under such a rescaling to regular Cartesian hydrodynamics. Somewhat counterintuitively, this leads to an effective cooling during a collapse when $\gamma < 5/3$. This also applies to the isothermal equation of state, commonly used in the study of turbulence in the interstellar medium and giant molecular clouds \citep{Robertson12,Hennebelle21}, albeit in terms of a time-dependent effective sound speed rather than an effective entropy production. 

In Appendix \ref{d dimensional conformal} we derive the conformal rescaling in an arbitrary number of dimensions, and show that the heating function vanishes for $\gamma$ corresponding to a monatomic gas in the relevant number of dimensions. This is because the gas particle velocity, $v_{\rm therm}$, undergoes the same rescaling as the gas fluid velocity under the (d-dimensional) isotropic rescaling, with $\langle v_{\rm therm}^2\rangle \propto |\mathbf{g}|^{-1/d}$ \citep{Martel98}. Writing the gas entropy in terms of the particle thermal velocity and gas density, 

\begin{align}
\begin{split}
 e^{s/R_{\rm gas}} &\propto \frac{T^{1/(\gamma - 1)}}{\rho} \propto \frac{\langle v_{\rm therm}^2 \rangle^{1/(\gamma - 1)}}{\rho} \\
&\propto |\mathbf{g}|^{\frac{\gamma - 1 - 2/d}{2 (\gamma - 1)}} \frac{\langle \tilde{v}_{\rm therm}^2 \rangle^{1/(\gamma - 1)}}{\tilde{\rho}} \propto  |\mathbf{g}|^{\frac{\gamma - 1 - 2/d}{2 (\gamma - 1)}}  e^{\tilde{s}/R_{\rm gas}}  .
\end{split}
\end{align}
We see that this rescaling is adiabatic when the gas is monatomic, with the right hand side of Equation \ref{entropy eq} vanishing. The original local-model equations are adiabatic, so the physical gas entropy $s$ must be conserved by the local flow. This means, for non-monatomic gases the rescaled entropy must evolve as $e^{\tilde{s}/R_{\rm gas}} \propto |\mathbf{g}|^{-\frac{\gamma - 1 - 2/d}{2 (\gamma - 1)}} $, in order that $e^{s/R_{\rm gas}}$ remain fixed, resulting in the source term in Equation \ref{entropy eq}. 

The rescaled vorticity for the conformal rescaling is given by $\tilde{\boldsymbol{\omega}} = (\boldsymbol{\partial} \times \boldsymbol{\gamma} \tilde{\mathbf{v}})$, this corresponds to the fluid vorticity of the rescaled problem as a result of $|\boldsymbol{\gamma}| = 1$. Substituting the expression for $p_0$ and $v_0$ into Equation \ref{vort ew} we recover the standard evolutionary equation for the vorticity, including baroclinic terms. For the apparent vorticity the isotropic source term is zero and its evolutionary equation becomes

\begin{align}
\begin{split}
 \tilde{\mathcal{D}} \boldsymbol{\omega}_{\mathbf{1}} &= \frac{1}{3} \tr (\boldsymbol{\gamma}^{-1}) \tilde{\rho}^{-2} (\boldsymbol{\partial} \tilde{\rho} \times \boldsymbol{\partial} \tilde{p}) \\
&- \boldsymbol{\partial} \times \left[ \boldsymbol{\gamma}^{-1} \boldsymbol{\gamma}_{\tau} \tilde{\textbf{v}} + \tilde{\rho}^{-1}  \left(\boldsymbol{\gamma}^{-1} - \frac{1}{3} \tr (\boldsymbol{\gamma}^{-1}) \mathbf{1} \right) \boldsymbol{\partial} \tilde{p}\right] 
\end{split} \quad .
\end{align}

As we shall show in Section \ref{incomp limit}, in general, the small scale/incompressible limit of these equations are independent of the heating function, meaning on small scales the hydrodynamics in conformally equivalent background flows are (asymptotically-)identical up to the conformal rescaling.

\subsection{2D rescaling}

For flows with $L_1 = L_2$ (how these correspond to directions in the global problem is left arbitrary\footnote{In the spherical flow setup directions 1 and 2 correspond to the horizontal directions with $L_1=L_2=\mathcal{R}$, while direction 3 corresponds to the radial direction with $L_3=L_z$.}), then another useful rescaling is possible by taking $v_0 = L_1^{-2}$ and $p_0 = L_1^{-4} L_3^{-1}$.  Considering 2 dimensional solutions to the local problem with $\tilde{v}_3 = \partial_3 = 0$, then the rescaled flow equations simplify to

\begin{align}
 \tilde{D} \tilde{\rho} &= - \tilde{\rho} \, \boldsymbol{\partial} \cdot \tilde{\mathbf{v}} , \label{rescale rho eq 2d} \\
 \tilde{\rho} \tilde{D} \tilde{v}^1 &= - \partial_1 \tilde{p} \\
 \tilde{\rho} \tilde{D} \tilde{v}^2 &= - \partial_2 \tilde{p} \\
\tilde{D} \tilde{p} + \gamma \tilde{p} \, \boldsymbol{\partial} \cdot \tilde{\mathbf{v}} &= \Gamma, \label{rescale p eq 2d}
\end{align}
where the $\tilde{v}_3$ momentum equation is automatically satisfied. The effective heating rate is given by

\begin{equation}
 \Gamma = \tilde{p}/t_{\rm heat} = -\left[ 2 (\gamma - 2) \frac{ \partial_{\tau} L_1}{L_1 } + (\gamma - 1) \frac{ \partial_{\tau}  L_3}{ L_3}   \right] \tilde{p} .
\end{equation}
For collapse profiles that satisfy

\begin{equation}
  L_3 \propto L_1^{-2 \frac{\gamma - 2}{\gamma - 1}}  , \label{2d condition}
\end{equation}
we recover regular (2D) ideal hydrodynamics. When holding $L_3$ fixed the system is fully 2-dimensionalised and we see that we require $\gamma=2$ to recover regular 2D hydrodynamics. As discussed in Appendix \ref{d dimensional conformal} this is a monatomic gas in 2-dimensions. For a 3D monatomic gas in the 2D rescaling the particle velocities are still distributed along the 3rd dimension, despite their mean $\tilde{v}^{3}=0$. This absorbs some of the energy from the horizontal acceleration, during a collapse, as a result of the equipartition of the thermal energy between the degrees of freedom, leading to a effective cooling.

\begin{figure}
\includegraphics[trim=20 10 50 20, clip, width=\linewidth]{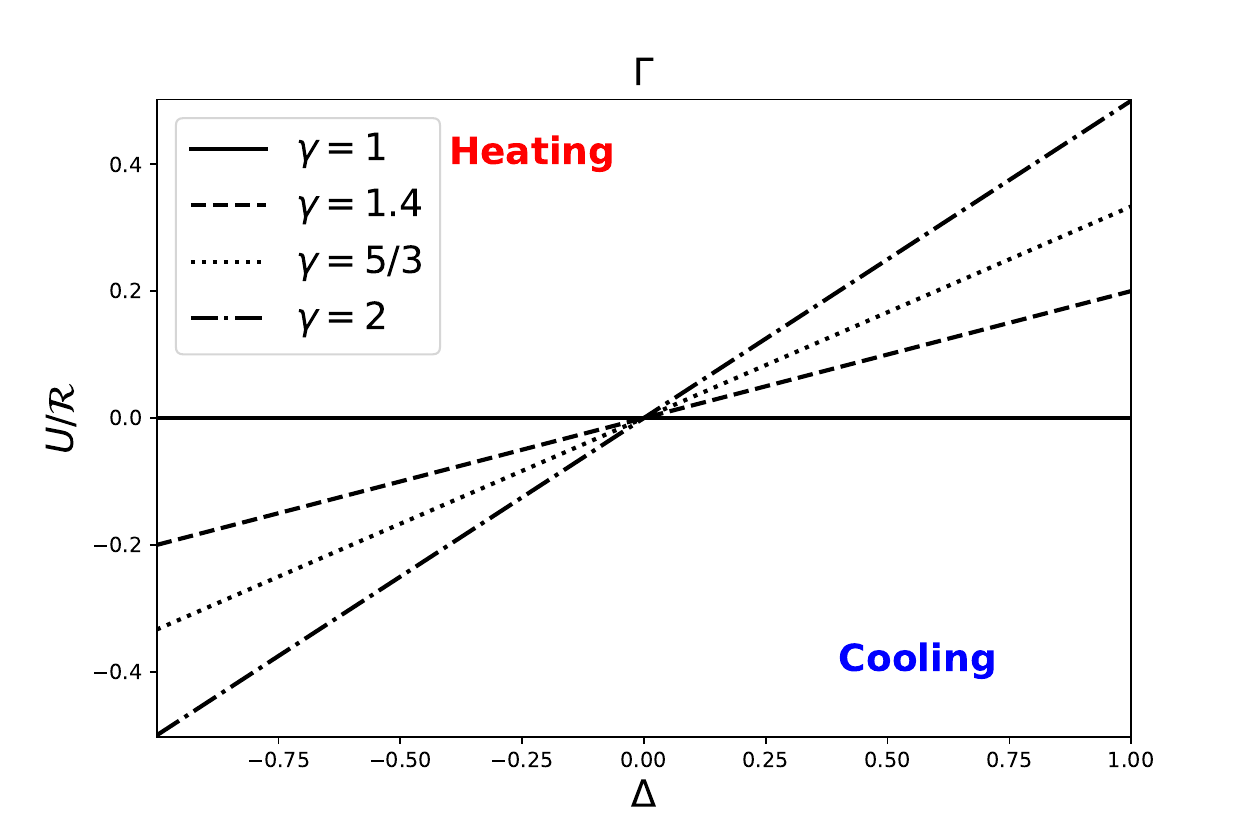}
\caption{Graph showing whether the background flow results in an effective heating or an effective cooling in the 2D rescaled problem. Here we are using the notation of \citet{Lynch23}, with $U/\mathcal{R} = \dot{L}_1/L_1$ and $\Delta$ being the velocity divergence of the background flow. The lines are the $\Gamma=0$ line for different ratio of specific heats. Spherical flows above these lines produce effective heating, while those below cool. }
\label{heating vs cooling}
\end{figure}

Figure \ref{heating vs cooling} shows how different background collapses lead to effective heating vs effective cooling in the rescaled problem. Notably there are situation in collapses (typically when $U/\mathcal{R} \ll \Delta$) where the collapse causes effective cooling in the rescaled problem (likewise there are similar situations in expansion where one has effective heating). This is particularly noticeable for $\gamma=1$ as the the rescaled problem always cools during a collapse and heats during expansion. 

The (rescaled) fluid vorticity for the 2D rescaling recovers the usual vorticity for a flow in 2D Euclidean space with $\tilde{\omega}^i = \hat{e}_{3}^i (\partial_1 \tilde{v}^{2} - \partial_2 \tilde{v}^{1})$, and the vorticity equation simplifies significantly to $\tilde{D} (\tilde{\omega}^3/\tilde{\rho}) = 0$.

\subsection{1D rescaling}

Finally for 1D solutions where (without further loss of generality) $\tilde{v}_1=\tilde{v}_2=\partial_1=\partial_2 = 0$, then we take $v_0 = L_3^{-2}$ and $p_0 = L_1^{-1} L_2^{-1} L_3^{-3}$, the rescaled fluid equations are

\begin{align}
 \tilde{D} \tilde{\rho} &= - \tilde{\rho} \, \partial_{3} \tilde{v}^{3} , \\
 \tilde{\rho} \tilde{D} \tilde{v}^3 &= - \partial_3 \tilde{p} , \\
\tilde{D} \tilde{p} + \gamma \tilde{p} \, \partial_{3} \tilde{v}^3 &= \Gamma, 
\end{align}
with the $\tilde{v}_1$ and $\tilde{v}_2$ components of the momentum equation automatically satisfied. The effective heating rate is given by

\begin{equation}
 \Gamma = \tilde{p}/t_{\rm heat} = -\left[ (\gamma-1) \left(\frac{\partial_{\tau} L_1}{L_1} + \frac{\partial_{\tau} L_2 }{L_2} \right)+(\gamma-3) \frac{ \partial_{\tau} L_3}{ L_3} \right] \tilde{p} .
\end{equation}
For collapse profiles that satisfy

\begin{equation}
 (L_1 L_2) \propto L_3^{-\frac{\gamma-3}{\gamma-1}} 
\end{equation}
we recover regular (1D) ideal hydrodynamics. Holding $L_1$ and $L_2$ fixed we require a 1D monatomic gas ($\gamma=3$) to obtain regular 1D hydrodynamics.

\section{Incompressible limit.} \label{incomp limit}

\subsection{Incompressible limit of the conformal rescaling} \label{conformal incomp}

In this section we derive the small scale/incompressible limit of the local model, i.e. we are looking to define the map $\Sigma : S \rightarrow S^{*}$ in Figure \ref{mapping cartoon}. This allows us to study small-scale/subsonic perturbations around this background flow using the extensive techniques developed to study incompressible fluid dynamics, similar to the use of incompressible shearing boxes in discs \citep[see ][for the equivalent derivation in discs]{Latter17}. While it may seem counterintuitive to take the incompressible limit of an evidently compressible flow such as a spherical collapse, the conformal rescaling introduced in Section \ref{conf rescl} absorbs the isotropic/volumetric part of the background flow. This means all such flows possess an incompressible limit, which can be understood as the composition of the maps in Figure \ref{mapping cartoon}. These limits are identical for conformally related background flows, which, unlike the compressible equations, applies regardless of the ratio of specific heats.

Starting with the conformally rescaled equations (Equations \ref{conformal cont}-\ref{conformal pre}) we split the rescaled quantities into a constant background state plus a nonlinear perturbation,

\begin{equation}
 \tilde{\mathbf{v}} = \mathbf{v}^{\prime} , \quad \tilde{\rho} = \bar{\rho} + \rho^{\prime} , \quad \tilde{p} = \bar{p} + p^{\prime} ,
\end{equation}
where $\bar{\rho}$ and $\bar{p}$ are the incompressible fluid density and background pressure respectively and primed quantities denote a (nonlinear) perturbation. We shall consider a situation where $\bar{\rho}$ is allowed to varying both spatially and temporally, but $\bar{p}$ is the spatially homogeneous background pressure. This will allow for density variations within the fluid, such as two separate incompressible fluid layers of differing densities. The resulting evolutionary equations for the perturbation are

\begin{align}
 \tilde{D} \bar{\rho} + \tilde{D} \rho^{\prime} + (\bar{\rho} + \rho^{\prime}) (\boldsymbol{\partial} \cdot \tilde{\mathbf{v}}) &= 0 ,\\
 \tilde{D} \tilde{\mathbf{v}}^{\prime}  + \boldsymbol{\gamma}^{-1} \boldsymbol{\gamma}_{\tau} \tilde{\mathbf{v}}^{\prime} &= - \frac{1}{\bar{\rho} + \rho^{\prime}} \boldsymbol{\gamma}^{-1} \boldsymbol{\partial} p^{\prime} , \\
 \tilde{D} \bar{p} + \tilde{D} p^{\prime} + \gamma (\bar{p} + p^{\prime} ) \left( \boldsymbol{\partial} \cdot \tilde{\mathbf{v}} \right) &= \frac{1}{2} \left( \gamma - \frac{5}{3}\right) \frac{\partial_{\tau} |\mathbf{g}|}{|\mathbf{g}|} (\bar{p} + p^{\prime})   .
\end{align}
Performing a nondimensionalisation of these equations, we introduce

\begin{align}
\begin{split}
\tilde{\mathbf{v}}^{\prime} &= w \mathbf{v}^{*} , \quad \mathbf{x} = \lambda \mathbf{x}^{*} , \quad \tau = (\lambda/w) \tau^{*} , \quad \bar{\rho} = \rho_{0} \rho^{*} , \\
\rho^{\prime} &=  \sigma \rho_{0} \rho^{*} \chi^{*} \quad p^{\prime} = \sigma \bar{p} \pi^{*}  ,  
\end{split}
\end{align}
where $*$'d variables denote nondimensionalised variables. We require $\rho^{*}$ be advected by the flow, without undergoing a change of volume, such that its Lagrangian derivative vanishes,

\begin{equation}
 \tilde{D} \rho^{*} = 0 .
\end{equation}
We also require the background pressure satisfy,

\begin{equation}
 \partial_{\tau^{*}} \bar{p} = \frac{1}{2} \left( \gamma - \frac{5}{3}\right) \frac{\partial_{\tau^{*}} |\mathbf{g}|}{|\mathbf{g}|} \bar{p}  ,
\end{equation}
i.e. $\bar{p}$ satisfies the evolutionary equation for the unperturbed background pressure of the rescaled problem. The nondimensionalised equations then become

\begin{align}
  (\boldsymbol{\partial}^{*} \cdot \mathbf{v}^{*}) &= - \sigma D^{*} \chi^{*} - \sigma \chi^{*} (\boldsymbol{\partial}^{*} \cdot \mathbf{v}^{*}) , \label{incomp general cont} \\
  D^{*}  \mathbf{v}^{*}  + \boldsymbol{\gamma}^{-1} \boldsymbol{\gamma}_{\tau^{*}} \mathbf{v}^{*} &= - \frac{\sigma \bar{p}}{\rho_0 w^2}\frac{1}{\rho^{*} (1 + \sigma \chi^{*})} \boldsymbol{\gamma}^{-1} \boldsymbol{\partial}^{*} \pi^{*} , \\
 \gamma \left( \boldsymbol{\partial}^{*} \cdot \mathbf{v}^{*} \right) &=  -\sigma D^{*} \pi^{*} - \gamma \sigma \pi^{*} \left( \boldsymbol{\partial}^{*} \cdot \mathbf{v}^{*} \right)  . \label{incomp general therm}
\end{align}
Setting $w^2 = \sigma \bar{p}_0/\rho_0$, where $\bar{p}_0 = \bar{p} (0)$ is the initial background pressure, then we can take the limit $\sigma \rightarrow 0$. This corresponds to taking the characteristic initial Mach number of the perturbations to zero, with $\sqrt{\sigma} = \frac{w}{\sqrt{p_0/\rho_0}} = \mathcal{M}_0$. Equations \ref{incomp general cont}-\ref{incomp general therm} then reduce to

\begin{align}
  (\boldsymbol{\partial}^{*} \cdot \mathbf{v}^{*}) &= 0 , \label{continuity old p} \\
  D^{*}  \mathbf{v}^{*}  + \boldsymbol{\gamma}^{-1} \boldsymbol{\gamma}_{\tau^{*}} \mathbf{v}^{*} &= -  \boldsymbol{\gamma}^{-1} \left( \frac{\boldsymbol{\partial}^{*} p^{*}}{\rho^{*}} \right) , \label{momentum old p}
\end{align}
where we have introduced $p^{*} = (\bar{p}/\bar{p}_0) \pi^{*}$. The thermal energy equation also recovers $(\boldsymbol{\partial}^{*} \cdot \mathbf{v}^{*}) = 0$. This limit $\sigma \rightarrow 0$, used to obtain the incompressible limit, is mostly independent of the validity of the local model. For example, the fluid motion can be incompressible with $\sigma \rightarrow 0$, but have significant rotational support, which is neglected by the local model.

The evolutionary equations for the rescaled vorticity are those from inviscid, incompressible hydrodynamics,

\begin{equation}
 D^{*} \boldsymbol{\omega}^{*} - \boldsymbol{\omega}^{*} \cdot \boldsymbol{\partial}^{*}  \mathbf{v}^{*} = (\rho^{*})^{-2} \boldsymbol{\partial} \rho^{*} \times \boldsymbol{\partial} p^{*} ,
\end{equation}
which contains a baroclinic source term in the presence of fluid layers of differing densities. The evolutionary equation for the apparent vorticity, $(\boldsymbol{\partial} \times \mathbf{v}^{*})$ is given by

\begin{align}
\begin{split}
 D^{*} \boldsymbol{\omega}^{*}_{\mathbf{1}} &- \boldsymbol{\omega}^{*}_{\mathbf{1}} \cdot \boldsymbol{\partial}^{*} \mathbf{v}^{*}  = \frac{1}{3} \tr (\gamma^{-1})  (\rho^{*})^{-2} (\boldsymbol{\partial}^{*} \rho^{*} \times \boldsymbol{\partial}^{*} p^{*} )  \\
&- \boldsymbol{\partial}^{*} \times \left[ \boldsymbol{\gamma}^{-1} \boldsymbol{\gamma}_{\tau^{*}} \textbf{v}^{*} + (\rho^{*})^{-1} \left(\boldsymbol{\gamma}^{-1} - \frac{1}{3} \tr (\gamma^{-1}) \mathbf{1} \right) \boldsymbol{\partial}^{*} p^{*}\right]  
\end{split} \quad ,
\end{align}
where the first source term is the baroclinic contribution and the second is from the flow anisotropy and there is no isotropic contribution.

For anisotropic collapses $\boldsymbol{\gamma}^{-1} \boldsymbol{\partial}^{*} p^{*}$ is not irrotational. Performing a Helmholtz decomposition of this vector,

\begin{equation}
\boldsymbol{\gamma}^{-1} \boldsymbol{\partial}^{*} p^{*} = \boldsymbol{\partial}^{*} P + \boldsymbol{\partial}^{*} \times \boldsymbol{\Omega} ,
\end{equation}
where $\boldsymbol{\Omega}$ has a gauge symmetry $\boldsymbol{\Omega} \rightarrow \boldsymbol{\Omega} + \boldsymbol{\partial} f$, for some scalar $f$, and we adopt a gauge in which $\boldsymbol{\partial}^{*} \cdot \boldsymbol{\Omega} = 0$. With this choice of gauge $\boldsymbol{\Omega}$ is related to $p^*$ by

\begin{equation}
 \Delta \boldsymbol{\Omega} = - \boldsymbol{\partial}^{*} \times ( \boldsymbol{\gamma}^{-1} \boldsymbol{\partial}^{*} p^{*} ) , \label{aniso pre Omega}
\end{equation}
thus $\boldsymbol{\Omega}$ is responsible for the anisotropic source term of the apparent vorticity. To enforce incompressibility we require

\begin{equation}
 \boldsymbol{\partial}^{*} \cdot [\mathbf{v}^{*} \cdot \partial \mathbf{v}^{*}  + \boldsymbol{\gamma}^{-1} \boldsymbol{\gamma}_{\tau^{*}} \mathbf{v}^{*} ] = - \rho^{*} \partial^{*T} \boldsymbol{\gamma}^{-1} \left(\frac{\boldsymbol{\partial} p^{*} }{\rho^{*}} \right) ,
\end{equation}
with the operator on the right hand side being related to the Laplacian operator with a metric $(\rho^{*})^2 \boldsymbol{\gamma}$. Inverting this equation for $p^{*}$ and substituting it into Equation \ref{aniso pre Omega}, it is possible to obtain an equation for $\boldsymbol{\Omega}$.

This procedure results in a new effective fluid pressure $P$ and incompressible fluid equations

\begin{align}
  (\boldsymbol{\partial}^{*} \cdot \mathbf{v}^{*}) &= 0 , \label{incomp continuity} \\
  \rho^{*} D^{*}  \mathbf{v}^{*}  + \boldsymbol{\partial}^{*} P  &= -\underbrace{( \rho^{*} \boldsymbol{\gamma}^{-1} \boldsymbol{\gamma}_{\tau^{*}} \mathbf{v}^{*}  + \boldsymbol{\partial}^{*} \times \boldsymbol{\Omega} )}_{\text{forces due to background anisotropy}}, \label{incomp momentum}
\end{align}
which have additional forces due to the anisotropy in the background flow. We thus see that, the small scale limit of a spherical flow can be mapped onto (inviscid) incompressible fluid dynamics, with additional forces due to anisotropy, upon a suitable rescaling. In the absence of anisotropy $\boldsymbol{\gamma} = \mathbf{1}$ and the effects of the collapse are entirely absorbed into the conformal rescaling; Equations \ref{incomp continuity}-\ref{incomp momentum} reduce to the usual equations for incompressible hydrodynamics.

Linear wave solutions to the anisotropic incompressible limit (Equations \ref{incomp continuity}-\ref{incomp momentum}) are derived in Appendix \ref{linear incomp}. Unlike with the addition of rotation or gravity, none of the vortical waves present in standard Cartesian incompressible fluids become oscillatory inertio-gravity modes. The solution does, however, exhibit the mixing property of the linear modes discussed in \citet{Lynch23}, where the two vortical waves source a third wave in the presence of anisotropy in the background flow. Instead of the vortical wave sourcing a (propagating) sound waves, as found in the compressible case, the vortical waves instead source a nonpropagating wave which is responsible for enforcing the incompressibility of the fluid motion in the presence of background anisotropy. This leads to a mixing of the velocities in the linear wave problem that can potentially lead to anisotropies in the fluid flow within the box starting from initially isotropic initial conditions.

\subsection{Effective Cosserat Media}

For the anisotropic fluid equation the final term is a form of odd elastic stress found in Cosserat media \citep{Fruchart23}. Cosserat media are a class of models in  continuum mechanics where every material point in the continuum is endowed with an intrinsic orientation that gives it additional rotational degrees of freedom. In models of solids this can represent the orientation of elongate crystals that make up the material, which will undergo rotations as the material deforms, with the material properties being dependent on the crystal orientation. As a result of these additional rotational degrees of freedom, each material point has an intrinsic angular momentum that is separate from the angular momentum of the fluid at that point.

The Cosserat Elastic force-stress tensor is \citep[see e.g.][]{Forest01}

\begin{equation}
 \sigma_{i j} = 2 \mu e_{(i j)} - 2 \mu_c e_{[i j]} ,
\end{equation}
where $\mu$ is the shear modulus and $\mu_c$ is the odd elasticity modulus, and we have introduce the Cosserat deformation tensor,

\begin{equation}
 e_{i j} = \partial_j \xi^*_i + \epsilon_{i j k} \varphi^k
\end{equation}
where $\xi_i$ and $\varphi_i$ are the displacement and rotation vectors. Choosing $\mu_c = - \mu$ then the fluid equations become

\begin{align}
  (\boldsymbol{\partial}^{*} \cdot \mathbf{v}^{*}) &= 0 , \\
 \rho^{*} D^{*}  \mathbf{v}^{*}  + \boldsymbol{\partial}^{*} P  &= -  \rho^{*} \boldsymbol{\gamma}^{-1} \boldsymbol{\gamma}_{\tau^{*}} \mathbf{v}^{*} + \boldsymbol{\partial}^{*} \cdot \boldsymbol{\sigma} , 
\end{align}
and we identify $\boldsymbol{\Omega} = 2 \mu \boldsymbol{\varphi}$, showing that $\boldsymbol{\Omega}$ is related to the micro-rotations of the effective Cosserat media. The rotation vector evolves according to the balance of the intrinsic angular momentum \citep[also referred to as moment of momentum][]{Forest01}, 

\begin{equation}
I \ddot{\boldsymbol{\varphi}} = \mathcal{F} (\rho^{*}, P, \boldsymbol{\partial}^{*} \mathbf{v}^{*}, \boldsymbol{\sigma}, \Delta \boldsymbol{\varphi}, \, \ldots \,) \quad ,
\end{equation}
where $I$ is a microphysical moment of inertia and the right hand side, $ \mathcal{F} $, depends on a combination of the force-stress tensor $\boldsymbol{\sigma}$, couple-stress tensor and volume couples \citep[see e.g.][for more details]{Forest01}. Based on Equation \ref{aniso pre Omega}, we see that the microphysical moment of inertia of the effective Cosserat media is zero. One can also show that the dominant balance in the moment of momentum equation must be between the couple-stress tensor divergence (responsible for terms involving $\Delta \boldsymbol{\varphi}$) and volume couples to produce the correct dynamics, however we shall not do that here. Considering Equation \ref{aniso pre Omega} and the equations for the apparent and true fluid vorticity we see that, in this reformulation, the Cosserat stress produces the anisotropic source term in the apparent vorticity equation that is responsible for enforcing conservation of the true fluid vorticity and accounts for the rotations performed by the background flow. 

\subsection{Incompressible limit of the 2D problem} \label{2d incomp}

If we instead consider the 2D rescaled equations (Equations \ref{rescale rho eq 2d}-\ref{rescale p eq 2d}) then we can similarly take the incompressible limit of these equations by introducing the same background+nonlinear perturbation, and nondimensionalisation, considered in Section \ref{conformal incomp}. Then taking $\sigma \ll 1$ we end up with the equations for a 2D incompressible fluid in a periodic domain. This shows that all the solutions of incompressible hydrodynamics on $\mathbb{T}^2$ are asymptotically equivalent to a subset of the solutions in any given spherical flow. 

\section{How rescaling time affects instabilities} \label{Rescaled time}

Hydrodynamical instabilities in the rescaled problem (for instance the Kelvin-Helmholtz instability that we will consider in Section \ref{KHI}) are exponentially growing in the rescaled time, $\tau$. This means they may not have sufficient time to grow in the physical time, $t$. before rotational support becomes important (which in many cases is likely to stabilise against the instability) or the finite rescaled time required for the instability to grow may correspond to an infinite physical time. If we consider a rescaled problem with a maximum reachable rescaled time of $\tau_{\rm max}$ then, in order for a hydrodynamical instability in the rescaled problem with growth rate $\sigma$ to be important in the spherical flow, we require the e-folding time of the instability to be less that $\tau_{\rm max}$, i.e. $\sigma > \tau_{\rm max}^{-1}$. This leads to a minimum growth rate for an instability in the rescaled problem to be relevant in the spherical flow of $\sigma_{\rm min} = \tau_{\rm max}^{-1}$.

In expanding flows the relative velocities tend to decrease with time, so the breakdown of the local model due to rotational support is not normally a problem. However, in many expanding flows it is not possible to reach all $\tau$ as it takes an infinite amount of physical time to reach a sufficiently large (but finite) $\tau$. If the e-folding time for a fluid instability in the rescaled problem is large enough then it requires an infinite physical time for such an instability to grow, meaning that instability doesn't operate in the expanding flow. Physically this corresponds to sufficiently slowly growing instabilities being geometrically attenuated (or frozen out, as used in the cosmology literature \citealt{Mo10}) by the background flow before they can grow to large amplitudes. The maximum reachable rescaled time, $\tau_{\rm max}$, for the expanding flow is

\begin{equation}
 \tau_{\rm max} = \int_0^{\infty} \frac{d t}{L^2}, 
\end{equation}
which can be finite for some expansion profiles. For instance in a Hubble like flow ($L = \exp (H t)$, for some constant Hubble parameter $H$) the maximum reachable rescaled time is

\begin{equation}
 \tau_{\rm max} = \frac{1}{2 H} .
\end{equation}
Again requiring a fluid instability undergo at least one e-folding then the minimum growth rate for an instability to operate in the expanding flow is

\begin{equation}
 \sigma_{\rm min} = \left( \int_0^{\infty} \frac{d t}{L^2} \right)^{-1} ,
\end{equation}
which in the specific example of the Hubble like flow is $ \sigma_{\rm min} = 2 H$.

For a collapse we must also consider at what point the growth in the horizontal flows lead to rotational support and a breakdown of the local model. For the full 3D (conformal) rescaling the conditions are given by Equations 25-27 of \citet{Lynch23}\footnote{Note there is a typo in Equation 27 which contains an extra factor of $R$ on the right-hand side of the inequality}. For the vertical (1D) case, the conditions for the validity of the local model are satisfied automatically as the horizontal flows are absent. Naively, for the horizontal (2D) case we would require \citep{Lynch23}

\begin{equation}
 \mathcal{R} (v^{x} v^{x} + v^{y} v^{y}) \ll  L_V^{-1} \min(c_s)^2 ,
\end{equation}
where $\mathcal{R}$ is the radial location of the box, $L_V$ is the vertical lengthscale associated with the box and $v^x$, $v^y$ are the horizontal velocities. However, this is predicated on there being vertical variation of fluid equations on the box lengthscale, which is not the case for 2D problems. These 2D solutions can still vary in the vertical/radial direction on the longer lengthscale $\sim R$. There are now two requirements for the 2D solutions to be valid. The first is for the rotation to not significantly affect the background flow which requires

\begin{equation}
 D_0 U + \partial_R \Phi + \frac{1}{\rho_0} \partial_{R} p_0 \gg R \langle v^x v^x + v^y v^y \rangle ,
\end{equation}
where the angle brackets denote some appropriate horizontal averaging over the box. This condition can be approximated by

\begin{equation}
 \mathcal{R}^{-3} \max(|\tilde{v}_{H}|)^2 \ll \max \left(\left| \frac{\partial_R p_0}{\rho_0} \right|, | U U_{R} |, |\partial_r \Phi |\right) , \label{rotational sup cond}
\end{equation}
where $\max(|\tilde{v}_{H}|)$ is the maximum horizontal velocity in the rescaled problem. Which term on the right hand side (pressure, advection, gravity) is important depends on the details of the background flow. The second way for the 2D problem to break down is a failure of the quasi-2D approximation due to vertical steepening of fluid quantities that vary on the longer radial lengthscale. This occurs on approximately the vertical nonlinear breaking time of the vertical velocity perturbation, which is

\begin{equation}
 t_{\rm break} \approx \frac{1}{\max (-\partial_R v^R)} \approx \left(\frac{v^R}{R}\right)^{-1} \approx \left(\frac{R (v^x v^x + v^y v^y) + \frac{p}{\rho R}}{R}\right)^{-1}  t_{\rm break}^{-1} ,
\end{equation}
where we have assumed $v^{R}$ is generated by a combination of the neglected rotational terms and vertical pressure gradients. We therefore require the physical time to satisfy

\begin{align}
\begin{split}
 t &< t_{\rm break} \\
&\approx \max \left[\mathcal{R}^{-2} \max(|\tilde{v}_{H}|), \mathcal{R}^{-1} R_0^{-1} \left (J/J_0 \right)^{-(\gamma - 1)/2} \max (c_{s 0})\right]^{-1} ,
\end{split}
\end{align}
with $\max (c_{s 0})$ the initial sound speed in the rescaled problem. In Appendix \ref{common flows} we determine the breakdown conditions for several example background flows.

\section{2D rescaled solutions} \label{2d sols}

In this section we demonstrate the utility of the 2D version of the rescaling transform to show how circular vortices and the Kelvin-Helmholtz instability can be found in spherical flows, using the solutions obtained in regular hydrodynamics.

\subsection{Isentropic, Circular, Vortex Solutions} \label{vortex sols}

In this section we look for the simplest vortex solutions to the local spherical flow model by utilising the rescaling symmetries. We consider vertically homogeneous solutions with $v^{3} = \partial_{3} = 0$ so that we can make use of the 2D rescaling discussed above (Equations \ref{rescale rho eq 2d}-\ref{rescale p eq 2d}). Consider a situation where Equation \ref{2d condition} is satisfied (for $\gamma\ne2$, for $\gamma=2$ we instead require $\dot{L}_1 = 0$) then $\Gamma=0$ and 2D solutions to the local spherical flow equations can be rescaled onto regular 2D hydrodynamics. As we are considering the 2D rescaling $L_1 = L_2$ and we identify $L_1 = L_2 = \mathcal{R}$ and $L_3 = L_{\tilde{z}}$ \citep[in the notation of][]{Lynch23} in spherical flows, and use $x$ and $y$ to denote coordinates along directions $\hat{e}^1$ and $\hat{e}^2$. We are looking for steady solutions to these equations which then become

\begin{align}
 \boldsymbol{\partial} \cdot (\tilde{\rho} \tilde{\mathbf{v}}) &= 0, \\
 \tilde{\rho} \tilde{\mathbf{v}}  \cdot \boldsymbol{\partial} v^1 &= -\partial_1 \tilde{p} , \\
 \tilde{\rho} \tilde{\mathbf{v}}  \cdot \boldsymbol{\partial} v^2 &= -\partial_2 \tilde{p} , \\
 \tilde{\mathbf{v}}  \cdot \boldsymbol{\partial} \tilde{p} &= - \gamma \tilde{p} \boldsymbol{\partial} \cdot \tilde{\mathbf{v}} .
\end{align}
We look for steady, circular, vortex solution to these equations. As the flow is inviscid it can exhibit a tangential discontinuity at the boundary of the vortex patch, which requires pressure continuity and no flows across the boundary. For a vortex of radius $r_{\rm vtx}$, we can obtain solutions of the form

\begin{align}
 \tilde{\rho} &= \tilde{\rho} (r) \Theta (r_{\rm vtx} - r)  + \tilde{\rho}_{\rm ext} \Theta (r - r_{\rm vtx})  , \\
 \tilde{\mathbf{v}} &= - \frac{1}{2} \mathcal{A} (r) \Theta (r_{\rm vtx} - r) (y \hat{\mathbf{e}}_x - x \hat{\mathbf{e}}_y) , \\
 \tilde{p} &= \tilde{p} (r) \Theta (r_{\rm vtx} - r)  + \tilde{p} (r_{\rm vtx}) \Theta (r - r_{\rm vtx}) ,
\end{align}
where $\tilde{\rho}_{\rm ext}$ is the uniform density of the fluid external to the vortex, $\Theta$ is the Heaviside step function and we have introduced the cylindrical radius from the vortex centre $r = \sqrt{x^2 + y^2}$. A constant $ \mathcal{A}$ corresponds to the vorticity of a uniform vorticity patch. These solutions work provided that

\begin{equation}
 \frac{d \tilde{p}}{d r} = \frac{1}{4} r  \mathcal{A} (r)^2 \tilde{\rho} . \label{vort pre cond}
\end{equation}
Taking $r_{\rm vtx} \rightarrow \infty$ recovers the isentropic Euler vortex commonly used to test numerical methods \citep{Shu98,Spiegel15}. As the 2D rescaled equations are invariant under translations and Galilean transformations, one can obtain uniformly translating vortices with an arbitrary initial location by applying the map $x \mapsto x - x_0 - v_c^x \tau$ and $y \mapsto y - y_0 - v_c^y \tau$, where $(x_0,y_0)$ are the initial position of the vortex and $(v_c^x,v_c^y)$ is the uniform velocity of the vortex centre, into the above expression.

Upon undoing the rescaling these vortex solutions become

\begin{align}
 \rho &= L_1^{-2} L_3^{-1} \left [ \tilde{\rho} (r) \Theta (r_{\rm vtx} - r)  + \tilde{\rho}_{\rm ext} \Theta (r - r_{\rm vtx})  \right], \\
 \mathbf{v} &= - \frac{1}{2}  \mathcal{A} (r) L_1^{-2} \Theta (r_{\rm vtx} - r)  [ (y - y_c) \hat{\mathbf{e}}_x - (x - x_c) \hat{\mathbf{e}}_y] , \\
 p &= L_1^{-4} L_3^{-1} \left[ \tilde{p} (r) \Theta (r_{\rm vtx} - r)  + \tilde{p} (r_{\rm vtx}) \Theta (r - r_{\rm vtx})  \right] ,
\end{align}
with $\tilde{\rho}$ and $\tilde{p}$ still related by Equation \ref{vort pre cond} and $r = \sqrt{(x-x_c)^2 + (y-y_c)^2}$. The vortex centre is located at 

\begin{equation}
(x_c,y_c) = \left(x_0 + v_c^x \int \frac{d t}{L_1^2} , y_0 + v_c^y \int \frac{d t}{L_1^2}  \right) ,
\end{equation}
which can be anticipated from the invariance of the local spherical flow under the modified Galilean transform \citep{Lynch23}. The vortex in the spherical flow has vorticity $\omega^{z} = L_1^{-2} \omega_0$ and we see that the vortex spins up as $L_1$ decreases. The vortex's translational velocity simultaneously speeds up.

\subsection{Kelvin-Helmholtz instability in spherical flows} \label{KHI}

In this section we use the rescaling transform to study how the Kelvin-Helmholtz instability \citep[][KHI]{vonHelmholtz1868,Thomson1871} operates in an contracting/expanding flow. As in Section \ref{vortex sols}, we consider vertically homogeneous solutions and make use of the incompressible limit of the 2D rescaling derived in Section \ref{2d incomp}. One can also find the compressible version of the KHI in the rescaled problem when $\Gamma=0$.

In the 2D-incompressible rescaled problem, consider two uniform density, uniform velocity planar flows separated by a vortex sheet in the y-direction. These planar flows having density and velocities $\tilde{\rho}_1$, $V_1$ and $\tilde{\rho}_2$, $V_2$ respectively. The dispersion relation for the perturbation to the shear layer interface can be obtained following the usual procedure \citep[e.g. see][]{Chandrasekhar81},

\begin{equation}
   \omega^2 + 2 \frac{\tilde{\rho}_1 V_1+ \tilde{\rho}_2 V_2}{\tilde{\rho}_1 + \tilde{\rho}_2} k_x \omega   + \frac{ \tilde{\rho}_1 V_1^2 +  \tilde{\rho}_2 V_2^2 }{\tilde{\rho}_1 + \tilde{\rho}_2} k_x^2 = 0 ,
\end{equation}
where $k_x$ is the wavenumber of the perturbation along the x-direction. From this we obtain

\begin{equation}
  \omega  =  -\frac{\tilde{\rho}_1 V_1 + \tilde{\rho}_2 V_2}{\tilde{\rho}_1 + \tilde{\rho}_2} k_x \pm i \sqrt{\tilde{\rho}_1 \tilde{\rho}_2 }\frac{|V_1 - V_2 | }{\tilde{\rho}_1 + \tilde{\rho}_2}   k_x 
\end{equation}
and we have the growth rate

\begin{equation}
 \sigma = \frac{\sqrt{\tilde{\rho}_1 \tilde{\rho}_2 }}{\tilde{\rho}_1 + \tilde{\rho}_2} |V_1 - V_2|   k_x . \label{KHI growth}
\end{equation}
For Hubble expansions this sets a maximum wavelength that is unstable to the Kelvin-Helmholtz instability as being

\begin{equation}
\lambda_{\rm max} = \frac{\sqrt{\tilde{\rho}_1 \tilde{\rho}_2 }}{\tilde{\rho}_1 + \tilde{\rho}_2} |V_1 - V_2|   \frac{\pi}{H} =  \frac{\sqrt{\rho_1 \rho_2 }}{\rho_1 + \rho_2} |v_1 - v_2| \frac{\pi L^2}{H} .
\end{equation}

\begin{figure}
\includegraphics[trim=0 0 50 20, clip, width=\linewidth]{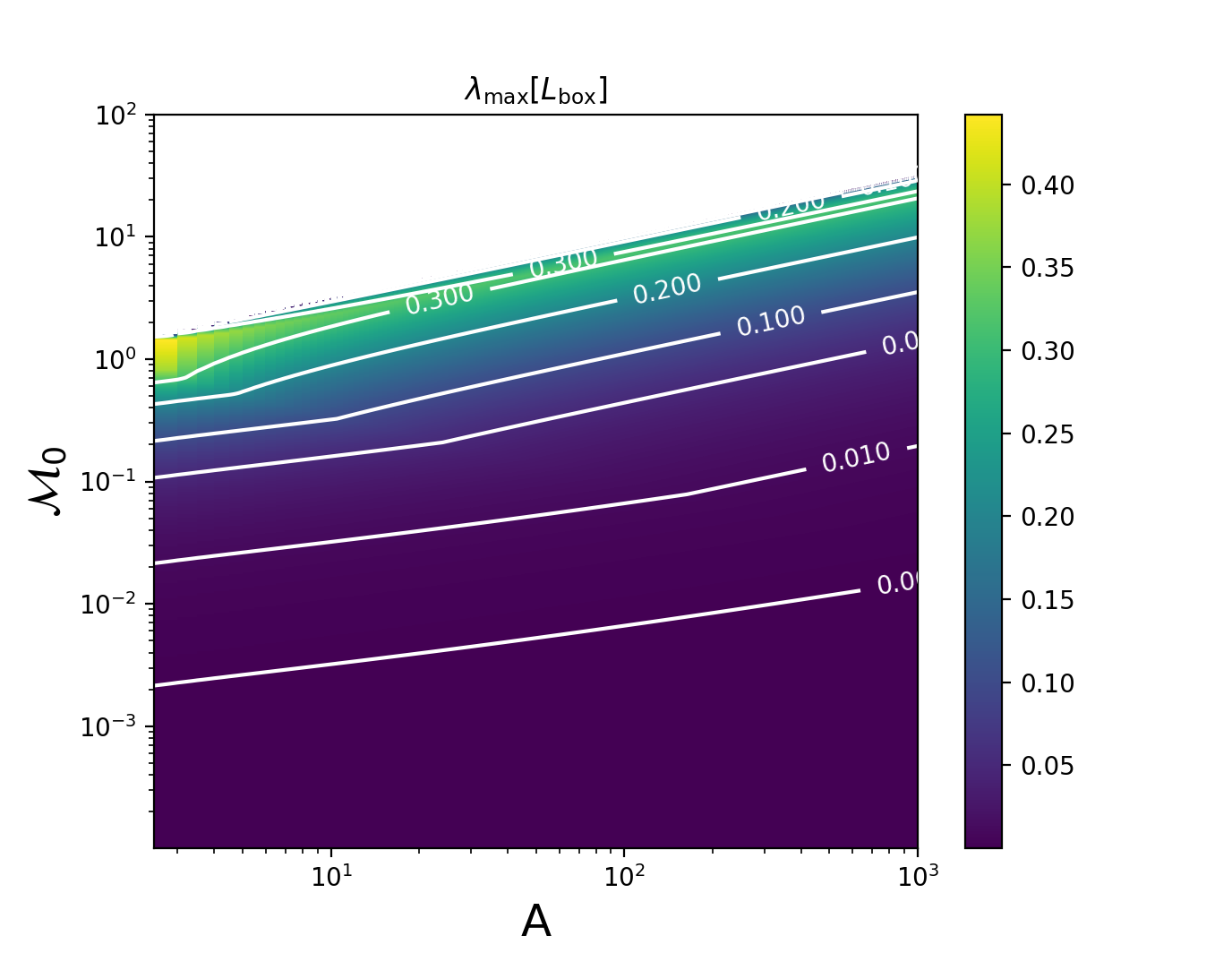}
\caption{Maximum unstable wavelength (in terms of the box-length) for the isothermal collapse model of \citet{Shu77}, for a uniform density medium, as a function of the initial Mach-number, $\mathcal{M}_0$, of the rotating band and the imbalance between the gravitational and pressure gradients in the background flow, $A$.}
\label{Maximum lambda Shu}
\end{figure}

Consider a rotating band on the equator of the isothermal, spherical collapse of \citet{Shu77}, with initial angular velocity $V$ and a uniform density throughout. Setting $V_1 = V R_0^2$ and $V_2 = 0$ (where we are implicitly using units in which $w=1$ and $c_s = O(\sigma^{-1/2})$ in the notation of Section \ref{incomp limit}), we are interested in calculating the maximum wavelength which can go unstable before the structure becomes rotationally supported, or the 2-dimensional approximation breaks down. The maximum rescaled time, $\tau_{\rm max}$ attainable in the \citet{Shu77} is given by Equation \ref{max tau shu} in Appendix \ref{shu model appendix}. By noting that the minimum unstable growth-rate satisfies $\sigma_{\rm min} \tau_{\rm max} = 1$ and making use of the Equation \ref{KHI growth}, for the growth rate, we obtain the maximum unstable wavelength to be 

\begin{equation}
\frac{\lambda_{\rm max}}{L_{\rm box}} = \frac{1}{2} \frac{\sqrt{\tilde{\rho}_1 \tilde{\rho}_2 }}{\tilde{\rho}_1 + \tilde{\rho}_2} \frac{\mathcal{M}_0}{ \sqrt{A - 2}} \ln\left|\frac{1 + \sqrt{1 - \left( \frac{R_{\rm min} (\mathcal{M}_0,A)}{R_0}\right)^2}}{1 - \sqrt{1 - \left( \frac{R_{\rm min}  (\mathcal{M}_0,A)}{R_0}\right)^2}} \right|  ,
\end{equation}
where $L_{\rm box}$ is the dimensionless box length, in the local model coordinates. Here $\mathcal{M}_0$ is the rescaled Mach number of the rotating band and $A$, taking the same meaning as in \citet{Shu77}, is a constant that determines the local imbalance between the gravitational and pressure gradients in the background flow. $R_{\rm min}$ is the minimum reachable radius by the spherical flow before the 2D local approximation breaks down. This is a function of $\mathcal{M}_0$ and $A$ that is determined in Appendix \ref{shu model appendix} Figure \ref{Maximum lambda Shu} shows how $\lambda_{\rm max}$ varies in the parameter space for the rotating band in a \citet{Shu77} collapse. The maximum unstable wavelength is maximised (at around the box lengthscale) for $\mathcal{M}_0 \rightarrow \sqrt{A}$, i.e. for the fastest rotating band where rotational support is still subdominant. However, as $\sqrt{A} > 1$ and we have used the incompressible equations, strictly $\mathcal{M}_0 \ll \sqrt{A}$ in the derivation of $\lambda_{\rm max}$. The modes with $\mathcal{M}_0 \sim \sqrt{A}$ would be expected to have significant corrections from compressibility effects and may be affected by the effective cooling $(\Gamma < 0)$ for the compressible, 2D, rescaling of the \citet{Shu77} collapse.

\begin{figure*}
\centering
\begin{subfigure}{0.4\textwidth}
\includegraphics[trim=50 50 105 20, clip, width=\textwidth]{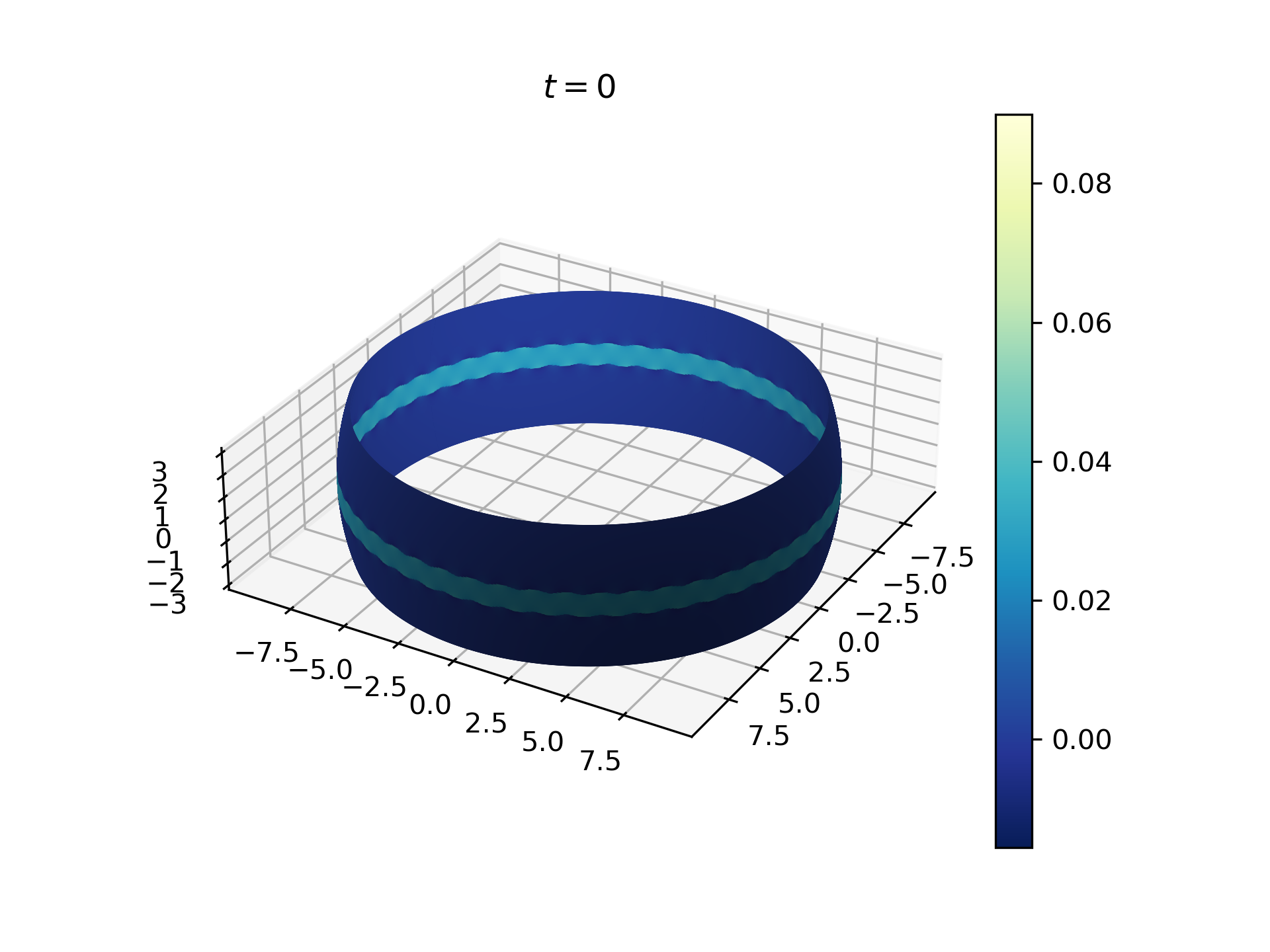}
\end{subfigure}
\begin{subfigure}{0.4\textwidth}
\includegraphics[trim=50 50 105 20, clip, width=\linewidth]{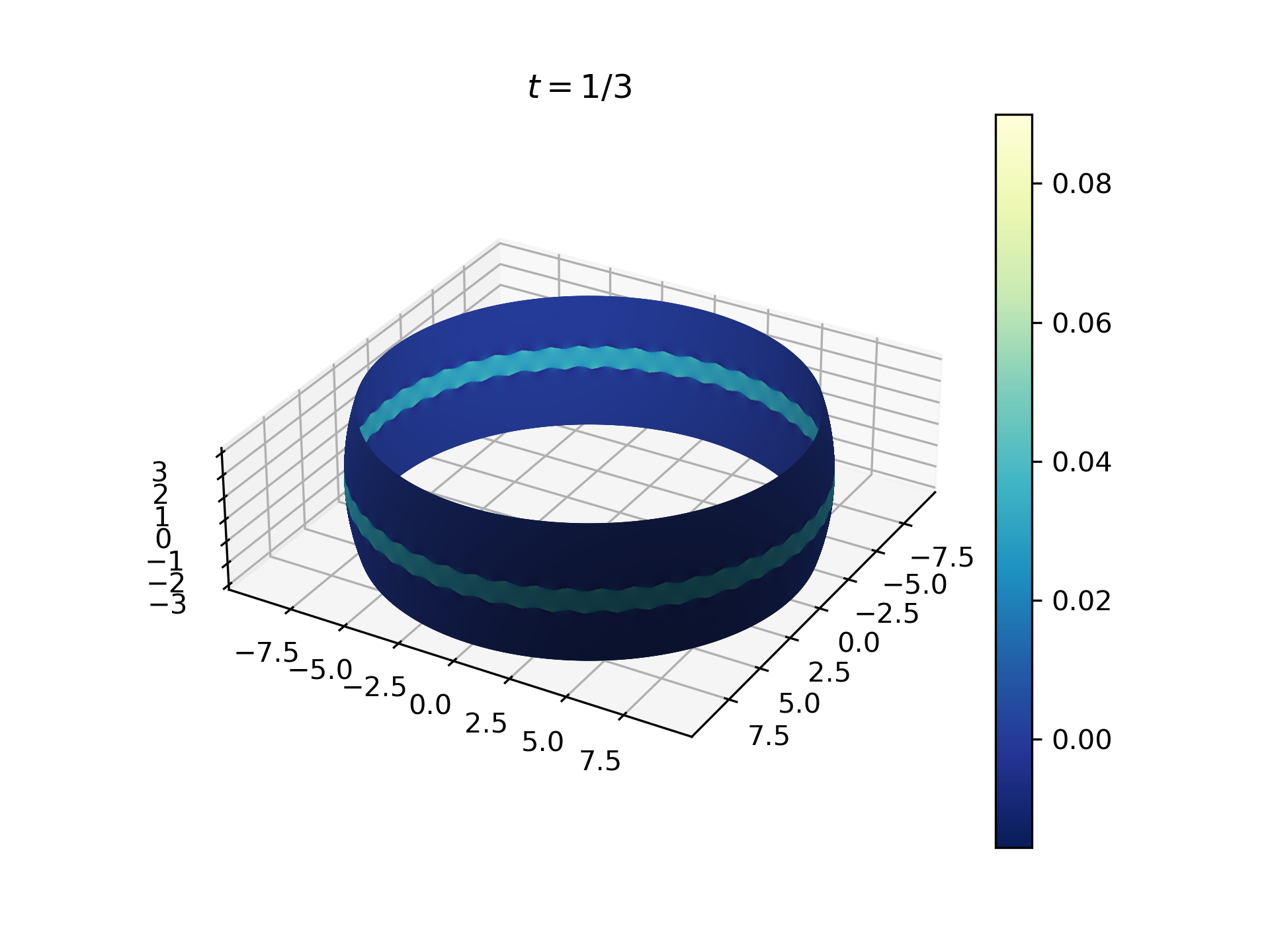}
\end{subfigure}
\begin{subfigure}{0.1\textwidth}
\hspace{3.0cm}
\end{subfigure}

\begin{subfigure}{0.4\textwidth}
\includegraphics[trim=50 50 105 20, clip, width=\linewidth]{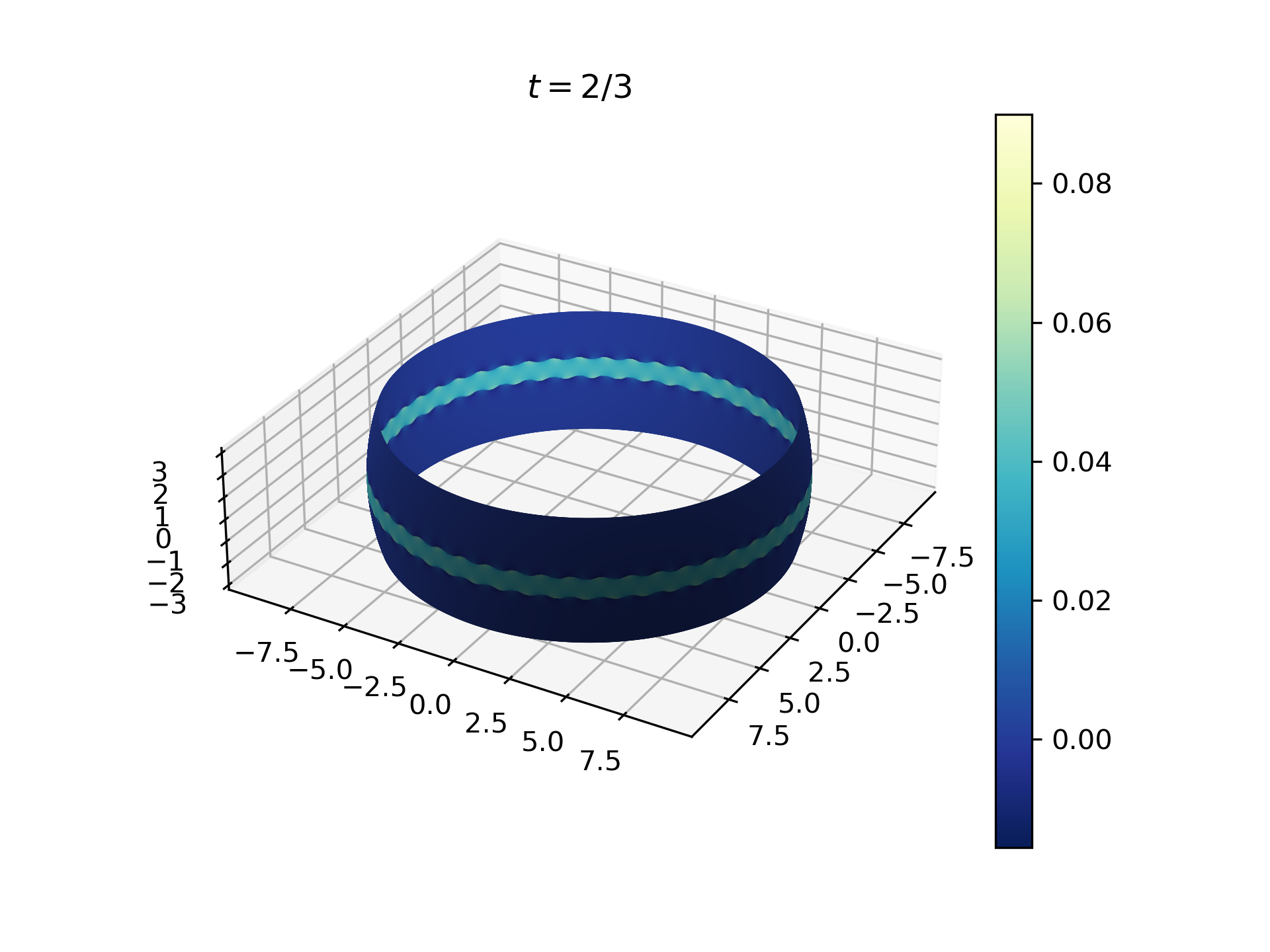}
\end{subfigure}
\begin{subfigure}{0.4\textwidth}
\includegraphics[trim=50 50 105 20, clip, width=\linewidth]{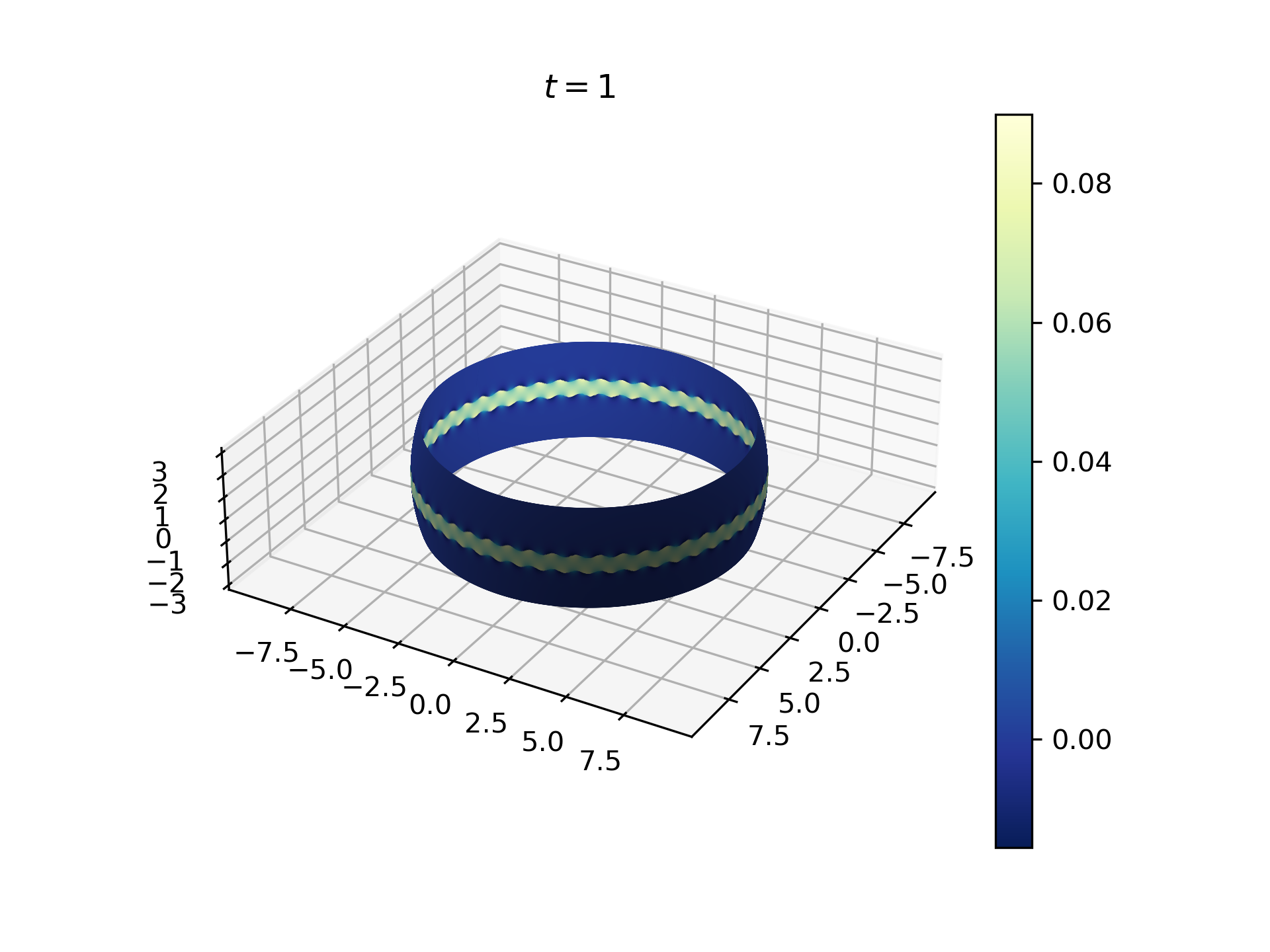}
\end{subfigure}
\begin{subfigure}{0.1\textwidth}
\includegraphics[trim=355 50 20 20, clip, width=\linewidth]{IMAGES/KHI_paper_0.png}
\end{subfigure}

\caption{Evolution of a rotating band on a spherical shell within the isothermal collapse of \citet{Shu77}, in the global picture. The parameters for this setup are $A = 3$, $a = 1$, $R_0 = 10$, $\mathcal{M}_0 = 0.3$, $k = 5$, with $10$ boxes around spaced around the midplane, chosen so that the unstable mode occurs on a large enough scale to be visible in the plot. Time is in units of the maximum timescale. As the collapse proceeds the rotating band spins up, while the KHI grows in amplitude (potentially enough to go nonlinear).  }
\label{KHI Shu Example}
\end{figure*}

\begin{figure*}
\centering
\begin{subfigure}{0.4\textwidth}
\includegraphics[trim=50 50 105 20, clip, width=\textwidth]{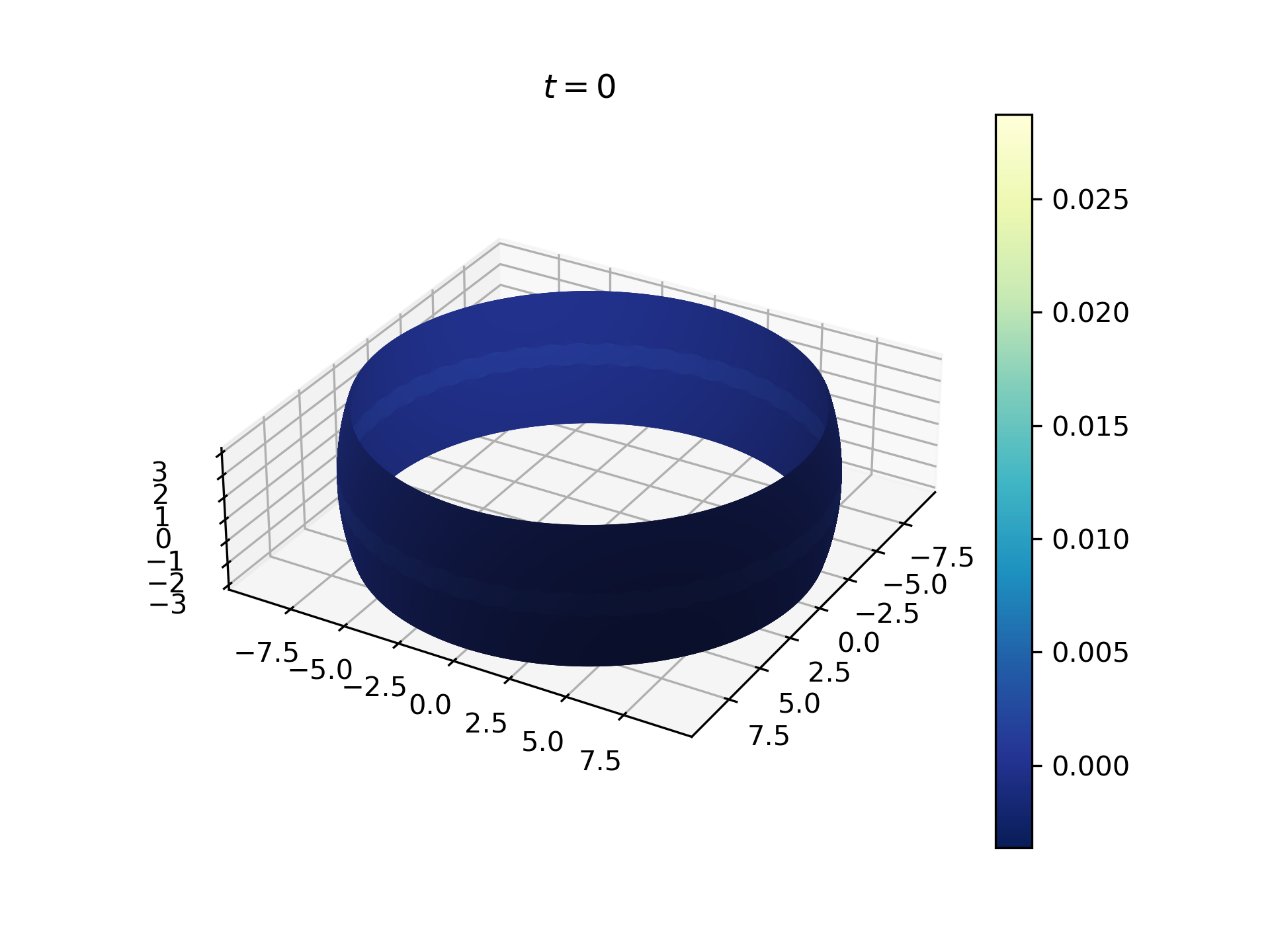}
\end{subfigure}
\begin{subfigure}{0.4\textwidth}
\includegraphics[trim=50 50 105 20, clip, width=\linewidth]{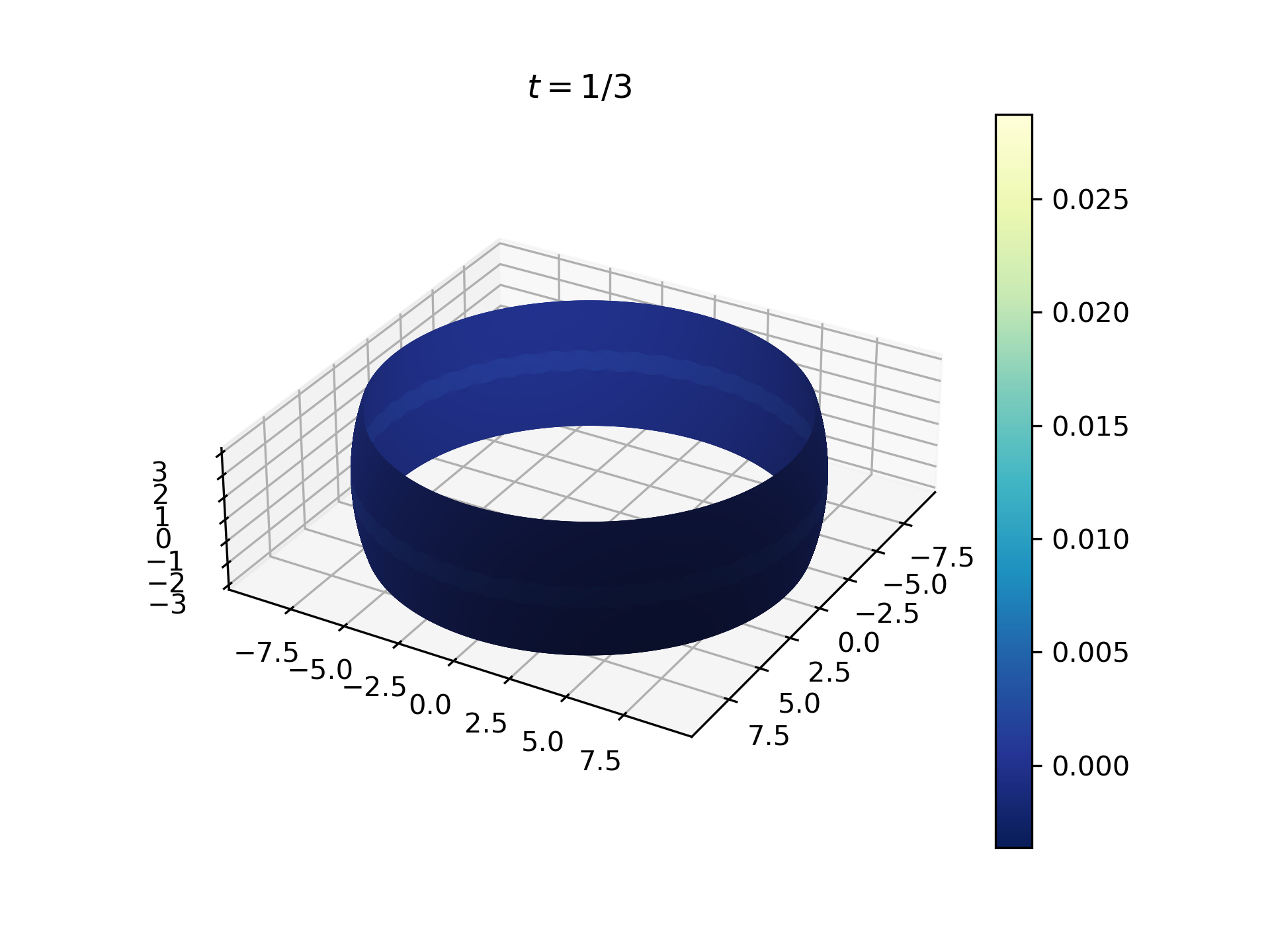}
\end{subfigure}
\begin{subfigure}{0.1\textwidth}
\hspace{3.0cm}
\end{subfigure}

\begin{subfigure}{0.4\textwidth}
\includegraphics[trim=50 50 105 20, clip, width=\linewidth]{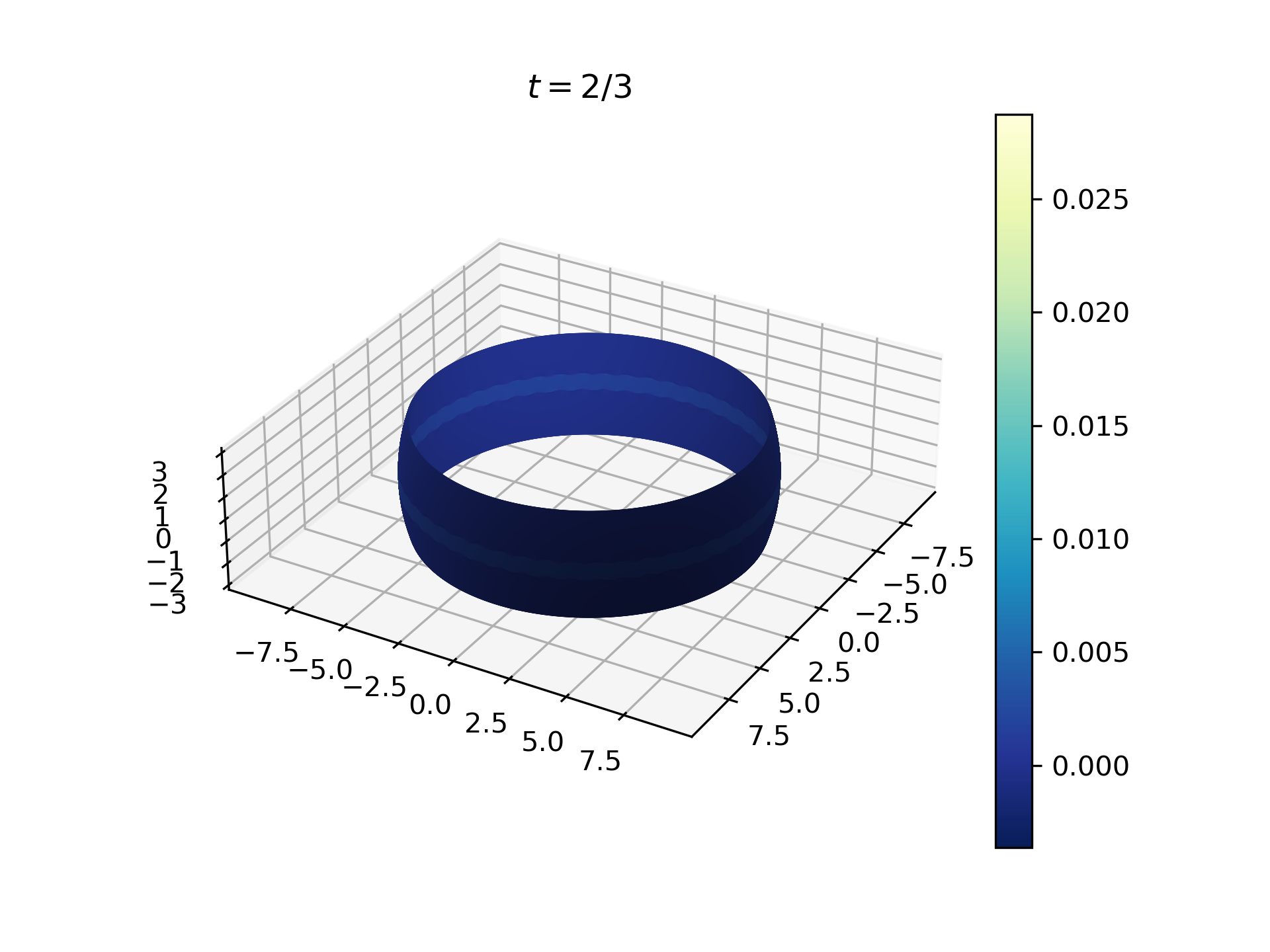}
\end{subfigure}
\begin{subfigure}{0.4\textwidth}
\includegraphics[trim=50 50 105 20, clip, width=\linewidth]{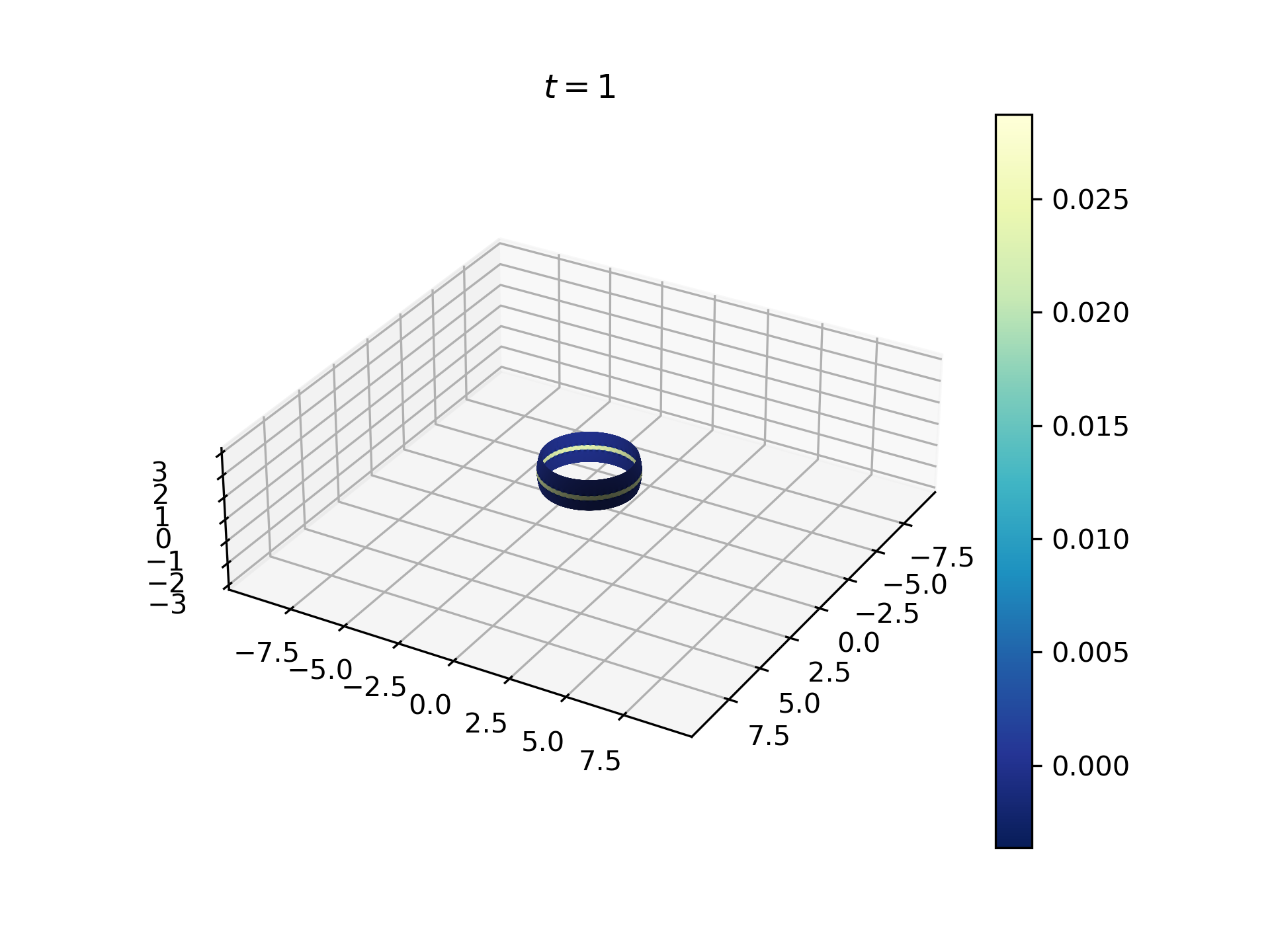}
\end{subfigure}
\begin{subfigure}{0.1\textwidth}
\includegraphics[trim=355 50 20 20, clip, width=\linewidth]{IMAGES/noKHI_paper_0.png}
\end{subfigure}

\caption{Same as Figure \ref{KHI Shu Example}, with $A=50$ and $\mathcal{M}_0 = 0.1$ but otherwise the same parameters as Figure \ref{KHI Shu Example}, resulting in no growth of the KHI on the collapse timescale.  }
\label{no KHI Shu Example}
\end{figure*}

Figure \ref{KHI Shu Example} shows an example evolution for the KHI for the rotating band on a \citet{Shu77} collapse. We have used the linearised solutions to the KHI \citep[e.g. from][]{Chandrasekhar81} here as this figure is primarily illustrative. A more in depth study of the KHI in rotating flows should be straightforward to carry out with the rescaling symmetries developed here - and can be done by rescaling the KHI test problems present in many astrophysical hydrocodes (e.g. present in \textsc{dedalus} for the incompressible case or \textsc{idefix} for the compressible). The different panels are equally spaced in physical time from the initial time to the maximum reachable time where the flow is expected to become 3-dimensional on the box lengthscale. The parameters for this setup are $A = 3$, $a = 1$, $R_0 = 10$, $\mathcal{M}_0 = 0.3$, $k = 5$, with $10$ boxes spaced around the midplane. These parameters don't strongly satisfy the asymptotics used to derive the model (particularly incompressibility) and are primarily chosen so that the instability has a long enough wavelength to be visible and has time to grow before the breakdown. More realistic parameters that better satisfy the assumptions of the (incompressible) local model will typically result in growth of the KHI on wavelengths much smaller than depicted here.  Figure \ref{no KHI Shu Example} shows a similar setup where the KHI is incapable of growing on the collapse timescale, for wavelengths chosen. Here we have $A=50$ and $\mathcal{M}_0 = 0.1$, with all other parameters the same as Figure \ref{KHI Shu Example}. Here the collapse proceeds much faster and the KHI does not have time to grow, however the rotating band spins up during the collapse until it reaches the point where vertical wavebreaking is expected to 3-dimensionalise the problem.

\section{Preliminary results on turbulence} \label{discuss}

In this work we have derived a set of rescalings for the fluid equations in the local spherical flow problem. We have used this to derive simple solutions/instabilities to flows within the local model, and show that this highly compressible flow still exhibits an incompressible limit that is relevant for fluid motion at small scales. In this section we shall discuss how the rescaling method could be used to study more complex fluid flows, which we will explore more thoroughly in future work. In particular the rescaling symmetries could be invaluable for the study of turbulence in spherical/Hubble flows.

The isotropic rescaling discussed here allows us to say more about the ``adiabatic heating'' discussed in \citet{Robertson12}. If one were to repeat their simulations with an adiabatic gas with $\gamma=5/3$, then the ``adiabatic heating'' effect is entirely absorbed into the rescaling. For the isothermal case considered by \citet{Robertson12}, however, the collapse in fact acts a form of cooling in the rescaled problem as the pressure grows slower than the density during the collapse. Only for an adiabatic gas with $\gamma>5/3$ does the collapse cause heating in the equivalent rescaled problem.

We now briefly outline the effects of an isotropic collapse on the turbulent power spectrum, using the rescaling properties. These results are valid for turbulence in adiabatic fluids with $\gamma=5/3$ or for turbulence on small scales where the fluid motion can be approximated as incompressible (where we expect Kolmogorov turbulence). In principle this might also apply in the high Mach number regime, where one expects Burgers turbulence, however this is less clear as Burger's equations are not a well defined limit of compressible hydrodynamics (discontinuities in Burger's equations having no relation to the strong shock limit of hydrodynamics). The box averaged kinetic energy is related to the power spectrum by

\begin{equation}
 \mathcal{E} = \left\langle \frac{1}{2} \rho \mathbf{v}^{T} \mathbf{g}\mathbf{v} \right\rangle= \int P_{k} \, d^d \mathbf{k} . 
\end{equation}
the wavenumber power spectrum is then

\begin{equation}
 P_{\mathbf{k}} (t) = \frac{1}{2} (\widehat{\sqrt{\rho} \mathbf{v} })^{\dagger}_{t \mathbf{k}} \mathbf{g} \, (\widehat{\sqrt{\rho} \mathbf{v}})_{t \mathbf{k}} ,
\end{equation}
where $\hat{\cdot}$ denotes a quantity in Fourier space and $\dagger$ denotes a conjugate-transpose. For the isotropic rescaling we have the following relationship between the physical and rescaled variables,

\begin{equation}
 (\widehat{\sqrt{\rho} \mathbf{v}})_{t \mathbf{k}} = |\mathbf{g}|^{-\frac{4+d}{4 d}} (  \widehat{\sqrt{\tilde{\rho}} \tilde{\mathbf{v}}})_{\tau(t) \mathbf{k}} ,
\end{equation}
resulting in the power spectrum transforming to

\begin{equation}
 P_{\mathbf{k}} (t) = |\mathbf{g}|^{-\frac{2+d}{2 d}} (  \widehat{\sqrt{\tilde{\rho}} \tilde{\mathbf{v}}})^{\dagger}_{\tau(t) \mathbf{k}} \boldsymbol{\gamma} \, (  \widehat{\sqrt{\tilde{\rho}} \tilde{\mathbf{v}}})_{\tau(t) \mathbf{k}} =  |\mathbf{g}|^{-\frac{2+d}{2 d}}  \tilde{P}_{\mathbf{k}} (\tau (t))
\end{equation}
where $\tilde{P}_{\mathbf{k}}$ is the power spectrum of the turbulence in the equivalent rescaled hydrodynamics and we see that the power spectrum undergoes the same rescaling as the pressure.

At small scales (and absent shocks) in an isotropically contracting/expanding flow the rescaling takes the contracting/expanding problem to regular, incompressible, 3D hydrodynamics, meaning we expect $\tilde{P}_{\mathbf{k}}$ to be set by decaying Kolmogorov turbulence. We thus see that the turbulence in isotropic contracting/expanding flows is a form of rescaled, decaying Kolmogorov turbulence (note the rescaling of time as well as the metric determinant scale factor). For collapses, where $\partial_{\tau} |\mathbf{g}| < 0$, it is possible for the ``decaying'' turbulence to increase in amplitude with time as a result of geometric enhancement.

For anisotropic flows one can also use the rescaling method to consider the effects of a rapid change in the box aspect ratio. Consider an initially isotropic box, exhibiting isotropic turbulence, that undergoes a rapid change in the box aspect ratio on a timescale much shorter than the turbulent correlation/eddy turnover time. In this limit the turbulence acts ``frozen in''. If we take a scale small enough such that the incompressible limit of the local model can be employed, and consider decaying Kolmogorov turbulence, then Equations \ref{vx lin}-\ref{vz lin} can be used to predict the turbulent velocity field, under a rapid change in aspect ratio, given an initially isotropic Kolmogorov spectrum for the velocity. We leave a detailed study of the effects of the collapse on turbulence (particularly the effects of anisotropy) for future work.

\section{Conclusion} \label{conc}

In this work we have developed an alternative approach to the spherical flow problem, based on the super-comoving variables introduced in the cosmological context, by demonstrating that the local hydrodynamics of various background flows are equivalent upon suitable rescaling. In this work we derived a set of conformal mappings which map local spherical flow problems into problems in Cartesian geometry with an effective heating term via a time-dependent rescaling of the fluid properties, metric tensor and time. 

\begin{itemize}
 \item A wide class of solutions to the local spherical flow model can be obtained via a mapping from the corresponding solution in regular Cartesian flows. We use this to obtain vortex solutions in the local model, that evolve due to the presence of the background flow.
 \item The effect of rescaling time constrains the instabilities that can occur, as instabilities present in the rescaled problem can be too slow to operate when mapped to the contracting/expanding problem. We show how this effects the linear phase of the Kelvin-Helmholtz instability in a spherical flow.
 \item On the small scale we show that spherical/Hubble flows have an incompressible limit, upon rescaling, with additional forces in the presence of anisotropy in the background flow.
\end{itemize}

In future work we intend to utilise the rescaling properties of the, incompressible limit of the spherical flow problem to demonstrate the universality of turbulence at small scales in spherical flows and explore the effects of anisotropy of the background flow on the turbulence, in particular its effect of generating anisotropy in the turbulence.

\section*{Acknowledgements}

This research was supported by the ERC through the CoG project PODCAST No 864965. This project has received funding from the European Union’s Horizon 2020 research and innovation programme under the Marie Skłodowska-Curie grant agreement No 823823. We thank Jeremy Fensch, Beno\^{i}t Commer\c{c}on, Ziyan Xu and Armand Leclerc for help with the literature. We thank the reviewer for suggestions that helped improve the readability of this paper.

\section*{Data availability}

No new data were generated or analysed in support of this research.



\bibliographystyle{mnras}
\bibliography{RescalingTransformLocalSphericalFlow.bib} 




\appendix

\section{Common flows and their rescaling} \label{common flows}

In this Appendix we derive the rescaled time, condition for $\Gamma=0$ for the set of flows considered in Appendix C of \citet{Lynch23} and compute the maximum reachable rescaled time during a collapse for the isothermal collapse model of \citet{Shu77}.

\subsection{Hubble like flows}

Consider a Hubble like flow with

\begin{equation}
 L = L_0 \exp(H t) ,
\end{equation}
where $H$ is a constant Hubble parameter.

For the conformal rescaling the heating function, $\Gamma$, is given by Equation \ref{heating func}. $\Gamma = 0$ when 

\begin{equation}
 L^2 J^{-(\gamma - 1)} = L^{5 - 3 \gamma} = \mathrm{const} \, ,
\end{equation}
meaning the conformal rescaling requires $\gamma=5/3$, as this corresponds to the super-comoving coordinates discussed in \citet{Shandarin80,Martel98}. As the background flow is isotropic the 2D and 1D transforms are identical.

The rescaled time for the Hubble like flow is

\begin{equation}
 \tau = \int_0^{t} \frac{d t}{L^2} = -\frac{1}{2 H L_0^2} [\exp(-2 H t) - 1 ] . 
\end{equation}

\subsection{Powerlaw flows}

Based on Appendix C of \citet{Lynch23} we have powerlaw flows with

\begin{equation}
 U = \mathcal{U}_0 (R/R_0)^{\beta} ,
\end{equation}
where the symbols have the same meaning as \citet{Lynch23}. For the powerlaw collapse, $\mathcal{R}$ and $L_z$ are given by

\begin{align}
 \mathcal{R} &= R_0 \left[1 + (1 - \beta) \mathcal{U}_0 R_0^{-1} t \right]^{1/(1 - \beta)} , \\
 L_z &= L_{z 0} \left[1 + (1 - \beta) \mathcal{U}_0 R_0^{-1} t \right]^{\beta/(1 - \beta)} ,
\end{align}
thus we have $L_z \propto \mathcal{R}^{\beta}$. 

For the conformal rescaling, $\Gamma=0$ for $\gamma=5/3$ as this is independent of the background flow for the conformal rescaling. The rescaled time is

\begin{equation}
 \tau  = - \frac{3 R_0}{(1 + 5 \beta) \mathcal{U}_0 J_0^{2/3}} \left[ (1 + (1 - \beta) \mathcal{U}_0 R_0^{-1} t )^{- ( 1 + 5 \beta)/3 (1 - \beta) }  - 1\right] .
\end{equation}
For the 2D rescaling, in order that $\Gamma=0$, we require 

\begin{equation}
  \mathcal{R}^2  (\mathcal{R}^2 L_z)^{-(\gamma - 1)} \propto \mathcal{R}^{- 2 (\gamma - 2) - (\gamma - 1) \beta  } = \mathrm{const} .
\end{equation}
This requires $\gamma = \frac{\beta + 4}{\beta + 2}$. An adiabatic index of $\gamma = 1.4$ requires $\beta = 3$. The associated rescaled timescale, $\tau$, is

\begin{equation}
 \tau  = -\frac{ 1 }{(1 + \beta ) \mathcal{U}_0 R_0} \left[ (1 + (1 - \beta) \mathcal{U}_0 R_0^{-1} t)^{-(1 + \beta)/(1 - \beta)} - 1 \right] .
\end{equation}
For the 1D rescaling we require

\begin{equation}
 L_z^2 (\mathcal{R}^2 L_z)^{-(\gamma - 1)} \propto \mathcal{R}^{(3 - \gamma) \beta - 2 (\gamma - 1)  } = \mathrm{const} ,
\end{equation}
in order that $\Gamma=0$. This, thus, requires $\gamma = \frac{2 + 3 \beta}{2 + \beta}$. For $\gamma = 1.4$, we require $\beta = 1/2$. The associated rescaled timescale, $\tau$ is

\begin{equation}
 \tau =  \frac{R_0}{(1 - 3 \beta) \mathcal{U}_0 L_{z 0}^2} \left[  (1 + (1 - \beta) \mathcal{U}_0 R_0^{-1} t)^{(3 \beta - 1)/(\beta - 1)} - 1 \right]  .
\end{equation}

\subsection{Isothermal collapse model of \citet{Shu77}} \label{shu model appendix}

For the isothermal spherical collapse problem of \citet{Shu77}, we have a background flow velocity of

\begin{equation}
 U = - a^2 (A - 2) t R^{-1} ,
\end{equation}
and $\mathcal{R}$ and $L_z$ are given by

\begin{align}
 \mathcal{R} &= R_0 \left[ 1 - \frac{a^2}{R_0^2} (A - 2) t^2 \right]^{1/2} , \\
 L_z &= L_{z 0} \left[ 1 - \frac{a^2}{R_0^2} (A - 2) t^2 \right]^{-1/2} ,
\end{align}
thus we have $L_z \propto \mathcal{R}^{-1}$.

This has $\Gamma=0$, with the conformal rescaling provided $\gamma=5/3$, which is not consistent with the isothermal equation of state assumed by the background flow. For both the 2D and 1D rescalings we again cannot have $\Gamma=0$, as this would requires a ratio of specific heats different from $1$. The rescaled time for the conformal rescaling is

\begin{equation}
 \tau = R_0^{-4/3} L_{z 0}^{-2/3}\int \left[ 1 - \frac{a^2}{R_0^2} (A - 2) t^2 \right]^{-1/3} d t,
\end{equation} 
while this integral can, in principle, be solved with hypergeometric functions it is more useful to obtain an (exact) series solution using a generalised binomial expansion

\begin{equation}
 \tau = \frac{1}{a \sqrt{A - 2}  L_{z 0}^{2/3} R_0^{1/3}}    \sum_{k=0}^{\infty} \begin{pmatrix}
-1/3 \\
k \end{pmatrix} \frac{(-1)^k}{2 k + 1} \left[ \frac{a}{R_0} \sqrt{A - 2} t \right]^{2 k + 1}   .
\end{equation} 
The associated rescaled time for the 2D rescaling is

\begin{equation}
 \tau = \frac{1}{2 R_0 a \sqrt{A - 2}} \ln\left|\frac{R_0 + a \sqrt{A - 2} t}{R_0 - a \sqrt{A - 2} t} \right| , \label{tau shu 2d}
\end{equation} 

\begin{figure}
\includegraphics[trim=0 0 60 20, clip, width=\linewidth]{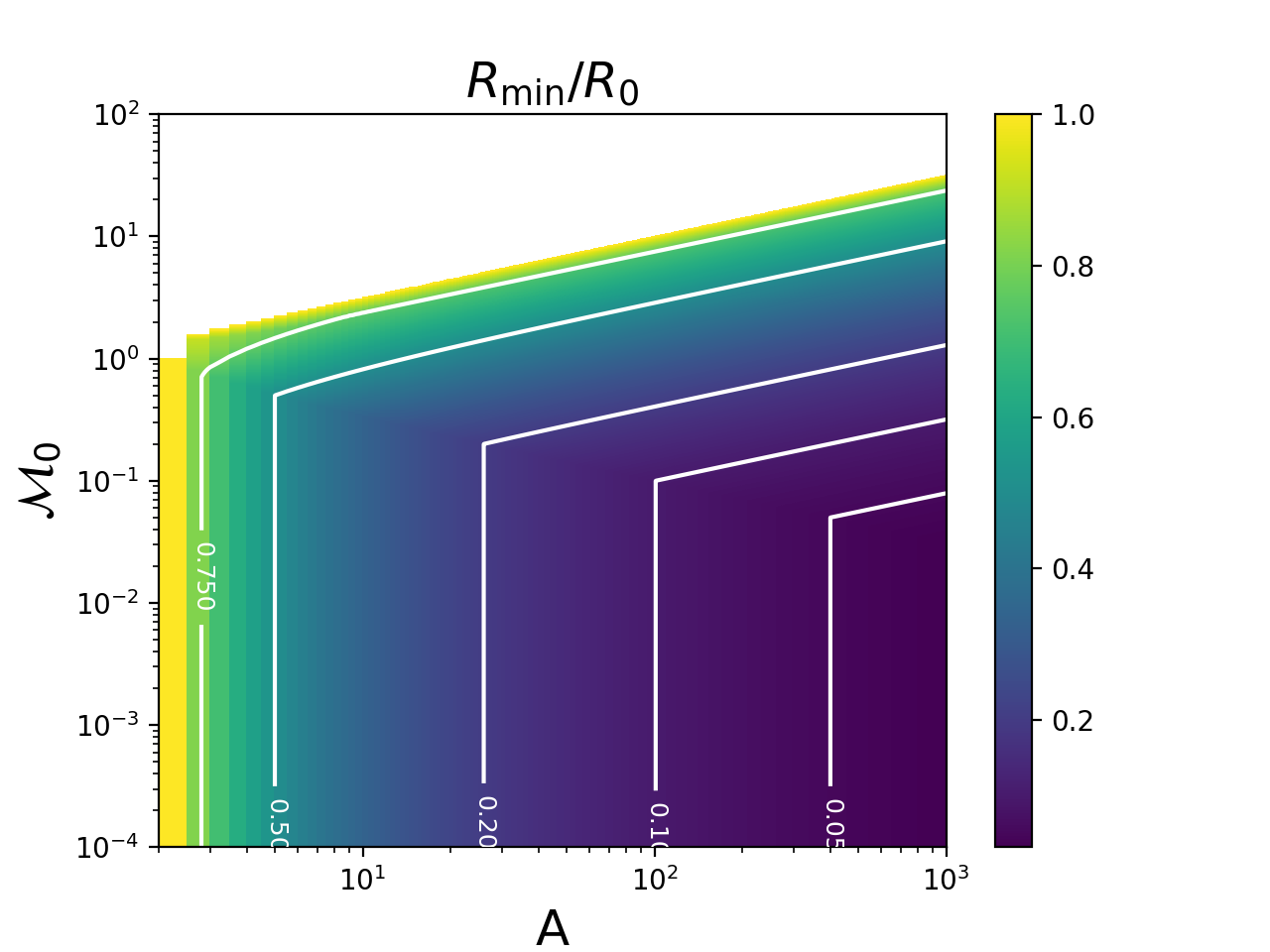}
\caption{Minimum reachable radius for the isothermal collapse model of \citet{Shu77}}
\label{Minimum radius Shu}
\end{figure}

In order to obtain the conditions for breakdown of the 2D solutions, the different terms on the left hand side of the inequality \ref{rotational sup cond} are

\begin{align}
 \left|\frac{\partial_{R} p_0}{\rho_0} \right| &= 2 \frac{a^2}{R} \quad , \\
 \left|U U_{R}\right| &= a^4 (A - 2)^2 t^2 R^{-3} \quad , \\
\left|\partial_{R} \Phi \right| &= A \frac{a^2}{R} \quad .
\end{align}
Making use of $A>2$ during a collapse the requirement to not have significant rotational support is

\begin{equation}
 \mathcal{M}_0^2 \ll \left ( \frac{\mathcal{R}}{R_0}\right)^{2}  \max\left(A , (A - 2)^2 \frac{a^2 t^2}{\mathcal{R}^2} \right) ,
\end{equation}
where we have introduced $ \mathcal{M}_0 = \max(|v_H|)/c_{s 0}$, the maximum initial Mach number in the rescaled problem. Writing the second term in terms of $\mathcal{R}$ we have

\begin{equation}
 \mathcal{M}_0^2 \ll  \max\left(A \left ( \frac{\mathcal{R}}{R_0}\right)^{2} , (A - 2) \left[1 - \left(\frac{\mathcal{R}}{R_0} \right)^2 \right] \right) . \label{rot sup cond}
\end{equation}
The right hand side has a minimum of $\frac{A}{2} \frac{A - 2}{A - 1}$, thus if 

\begin{equation}
  \mathcal{M}_0^2 < \frac{A}{2} \frac{A - 2}{A - 1} ,
\end{equation}
then the local flow will not become rotationally supported before reaching $R=0$. As the first term in condition \ref{rot sup cond} is largest at large $\mathcal{R}$, then this condition will be violated first when $\mathcal{M}_0^2 = A \left( \mathcal{R}/R_0\right)^2$, allowing us to determine the radius at which the local model breaks down due to rotational support,

\begin{equation}
\frac{R_{\rm rot}}{R_0} = \frac{\mathcal{M}_0}{\sqrt{A}} .
\end{equation}
The nonlinear breaking time (or timescale for the problem to become 3-dimensional) for this problem can be expressed as

\begin{equation}
 \frac{a^2 t_{\rm break}^2}{R^2} = \max \left(1, \left(\frac{\mathcal{R}}{R_0} \right)^{-2} \mathcal{M}_0^2 \right)^{-1} \quad. \label{breaking time}
\end{equation}
Introducing the radius at which the problem becomes 3-dimensional $R_{\rm 3D} = \mathcal{R} (t_{\rm break})$, with $\mathcal{R} \ge R_{\rm 3D}$ then if

\begin{equation}
 \mathcal{M}_{0} \le \frac{R_{\rm 3D}}{R_0} , 
\end{equation}
then the first term on the left hand side of Equation \ref{breaking time} is dominant and we obtain for the 3-dimensionalising radius,

\begin{equation}
 \frac{R_{\rm 3D}}{R_0} = \frac{1}{\sqrt{A - 1}} .
\end{equation}
This allows us to determine the Mach number cutoff to be $\mathcal{M}_{0} \le \frac{1}{\sqrt{A - 1}}$. For $\mathcal{M}_{0} > \frac{1}{\sqrt{A - 1}}$ the second term on the right-hand side of Equation \ref{breaking time} dominates and, after selecting the positive real root, we obtain the following for the 3-dimensionalising radius,

\begin{equation}
 \frac{R_{\rm 3D}}{R_0} = \frac{\mathcal{M}_0}{\sqrt{A - 2}} \sqrt{-\frac{1}{2} + \left( \frac{1}{4} + \frac{A - 2}{\mathcal{M}_0^2}\right)^{1/2}} \quad .
\end{equation}
The minimum reachable radius, $R_{\rm min}$, is then the first of $R_{\rm 3D}$ or $R_{\rm rot}$ to be reached by the flow, i.e. 

\begin{equation}
 R_{\rm min} = \max (R_{\rm 3D} , R_{\rm rot} ).
\end{equation}
If $R_{\rm min} \ge R_0$ then the local model is not valid for the chosen initial conditions. The minimum reachable radius for the \citet{Shu77} model is a function of the initial Mach number of the perturbations in the rescaled flow and $A$ and is plotted in Figure \ref{Minimum radius Shu}. As can be seen to allow the 2-dimensional local solutions to be good approximations to small scale perturbations of the global problem we require small $\mathcal{M}_0$ and large $A$, i.e. initially highly subsonic perturbations to flows that are close to free-fall. 

\begin{figure}
\includegraphics[trim=0 0 60 20, clip, width=\linewidth]{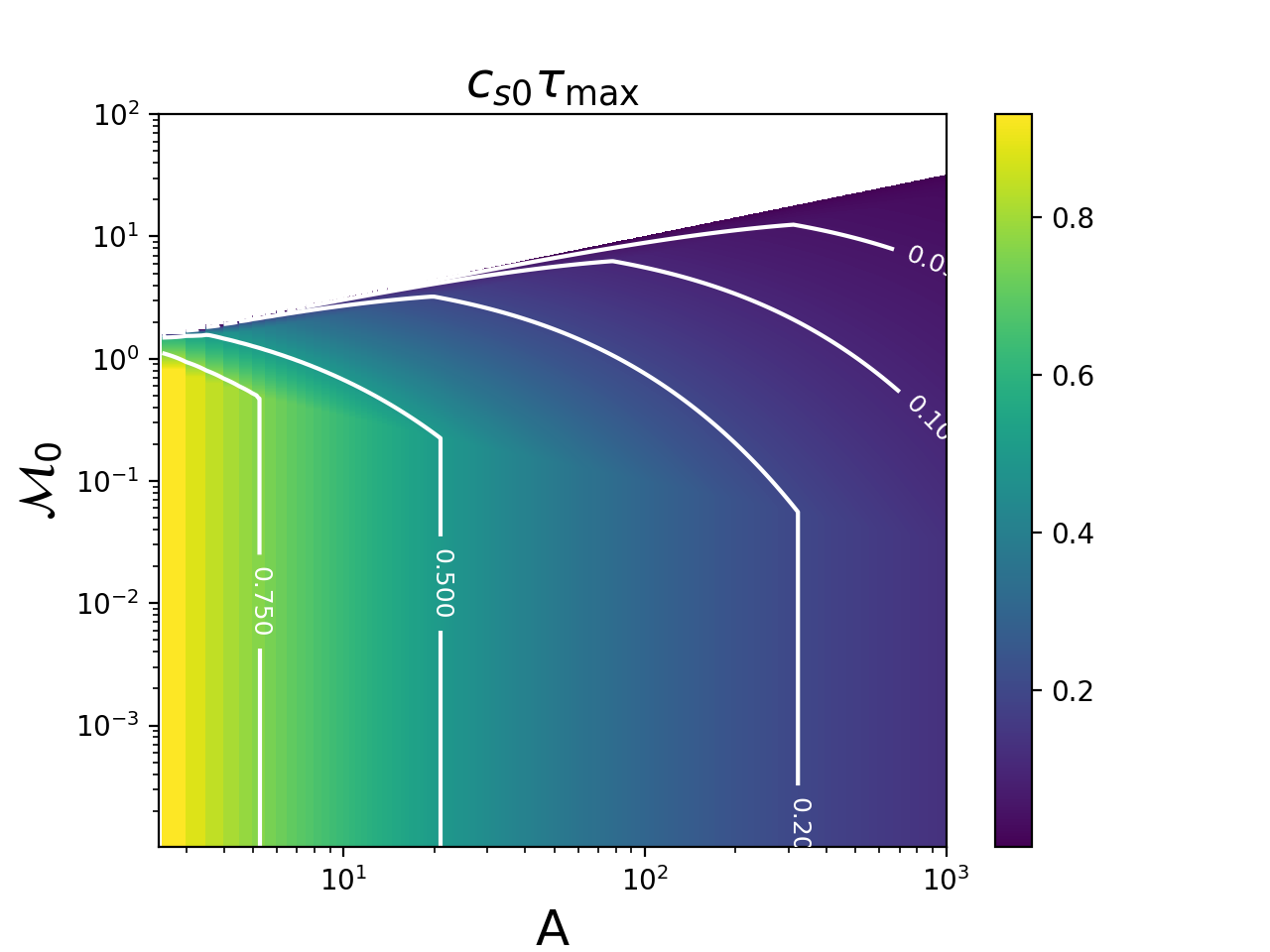}
\caption{Maximum reachable rescaled time, during a collapse, for the isothermal collapse model of \citet{Shu77}}
\label{Maximum tau Shu}
\end{figure}

The corresponding maximum reachable rescaled time can be obtained by substituting in $t (R_{\rm min})$ into equation \ref{tau shu 2d}, 

\begin{equation}
 \tau_{\rm max} = \frac{1}{2 R_0 a \sqrt{A - 2}} \ln\left|\frac{1 + \sqrt{1 - \left( \frac{R_{\rm min}}{R_0}\right)^2}}{1 - \sqrt{1 - \left( \frac{R_{\rm min}}{R_0}\right)^2}} \right| , \label{max tau shu}
\end{equation} 
which is important for determining what quasi-2-dimensional instabilities can operate in the collapse. This is plotted in Figure \ref{Maximum tau Shu}. Here we see that to attain large rescaled time we instead need small $A$ (and $\mathcal{M}_0$), i.e. subsonic perturbations to slowly contracting pressure supported structures, principally due to the factor of $\sqrt{A - 2}$ in front of the time in the equation for the radius. In the limit of arbitrarily small $\mathcal{M}_0$ the maximum reachable rescaled time becomes

\begin{equation}
 \lim_{\mathcal{M}_0 \rightarrow 0} \tau_{\rm max} = \frac{1}{2 R_0 a \sqrt{A - 2}} \ln\left|\frac{\sqrt{A - 1} + \sqrt{A-2}}{\sqrt{A - 1} - \sqrt{A-2} } \right| .
\end{equation} 
For subsonic background flows with $A = 2 + \epsilon$ this becomes

\begin{equation}
 \lim_{\mathcal{M}_0 \rightarrow 0} \tau_{\rm max} = \frac{1}{R_0 a} (1 - \epsilon/6) +  O (\epsilon^{3/2}) .
\end{equation} 
For background flows close to freefall ($A \gg 1$), this is instead

\begin{equation}
 \lim_{\mathcal{M}_0 \rightarrow 0} \tau_{\rm max} = \frac{1}{2 R_0 a} A^{-1/2} \ln 4 A  + O (A^{-3/2} \log A) .
\end{equation} 

Finally for the 1D rescaling, the associated rescaled time is

\begin{equation}
 \tau = L_{z 0}^{-2} t \left[ 1 - \frac{a^2}{3 R_0^2} (A - 2) t^2 \right]  .
\end{equation}

\section{Evolution of the apparent vorticity} \label{apprent vort equation}

In this appendix we consider the evolution of the ``apparent'' vorticity $\boldsymbol{\omega}_{\mathbf{1}} = \boldsymbol{\partial} \times \tilde{\mathbf{v}}$, which corresponds to the vorticity of the analogue problem in the Cartesian box. This evolves according to

\begin{equation}
 \tilde{D} \boldsymbol{\omega}_{\mathbf{1}} - \boldsymbol{\omega}_{\mathbf{1}} \cdot \boldsymbol{\partial} \tilde{\mathbf{v}} + \boldsymbol{\omega}_{\mathbf{1}} \boldsymbol{\partial} \cdot \tilde{\mathbf{v}} = -  \frac{\dot{v}_{0}}{v_0^2} \boldsymbol{\omega}_{\mathbf{1}} - \boldsymbol{\partial} \times \left[ \mathbf{g}^{-1} \dot{\mathbf{g}} \tilde{\textbf{v}} +  \frac{p_0 |\mathbf{g}|^{1/2}}{v_0^2 \tilde{\rho}} \mathbf{g}^{-1} \boldsymbol{\partial} \tilde{p}\right] . \label{apparent vort eq}
\end{equation}

We can separate out the source terms into contributions from baroclinicity, the isotropic rescaling/volumetric part of the background flow, and flow anisotropy as follows

\begin{align}
\begin{split}
 \tilde{\mathcal{D}} \boldsymbol{\omega}_{\mathbf{1}} &= - v_0^{-1} \underbrace{\left( \frac{\dot{v}_{0}}{v_0} + \frac{2}{3} \Delta \right) \boldsymbol{\omega}_{\mathbf{1}}}_{\text{isotropic}} + \frac{1}{3} \underbrace{ \tr (\mathbf{g}^{-1}) \frac{p_0 |\mathbf{g}|^{1/2}}{v_0^2 \tilde{\rho}^2} (\boldsymbol{\partial} \tilde{\rho} \times \boldsymbol{\partial} \tilde{p})}_{\text{baroclinic}} \\
&- \underbrace{\boldsymbol{\partial} \times \left[ v_0^{-1} \left(\mathbf{g}^{-1} \dot{\mathbf{g}} - \frac{2}{3} \Delta \mathbf{1} \right) \tilde{\textbf{v}} + \frac{p_0 |\mathbf{g}|^{1/2}}{v_0^2 \tilde{\rho}}  \left(\mathbf{g}^{-1} - \frac{1}{3} \tr (\mathbf{g}^{-1}) \mathbf{1} \right) \boldsymbol{\partial} \tilde{p}\right]  }_{\text{anisotropy}}
\end{split} \quad ,
\end{align}
where $\tilde{\mathcal{D}}$ is an advection operator (a weighted Lie-derivative) that, when acting on $\boldsymbol{\omega}_{\mathbf{1}}$, yields the left hand side of Equation \ref{apparent vort eq} . The isotropic terms appear depending on the choice of rescaling, and arise if the rescaling is not chosen to absorb the isotropic part of the background flow. The anisotropic terms are responsible for enforcing conservation of the true fluid vorticity $\tilde{\omega}$, which contains a contribution from the metric tensor. The baroclinic source terms appear due to being a source of the true fluid vorticity.

\section{Conformal rescaling in d-dimensions.} \label{d dimensional conformal}

For an ideal fluid in d-dimensions the local spherical flow equations can be generalised to

\begin{align}
 D \rho = - \rho \left[\Delta + \boldsymbol{\partial} \cdot \mathbf{v} \right] \\
 \rho D \mathbf{v} + \rho \mathbf{g}^{-1} \dot{\mathbf{g}} \mathbf{v} = - \mathbf{g}^{-1} \boldsymbol{\partial} p , \\
 D p = - \gamma p \left[\Delta + \boldsymbol{\partial} \cdot \mathbf{v} \right] ,
\end{align}
where $D = \partial_t + \mathbf{v} \cdot \boldsymbol{\partial}$ is the Lagrangian time derivative with respect to the relative flow, $\Delta$ is the velocity divergence of the background flow and $\mathbf{g}$ is a diagonal metric tensor,

\begin{equation} 
 \mathbf{g} = \diag (L_1^2, \cdots, L_d^2) .
\end{equation}
Introduce the conformal metric, $\boldsymbol{\gamma}$, related to the original metric by  

\begin{equation}
 \mathbf{g} = |\mathbf{g}|^{1/d} \boldsymbol{\gamma} , 
\end{equation}
and a time-dependent rescaling of the variables,

\begin{equation}
 \rho =  |\mathbf{g}|^{-1/2} \tilde{\rho} , \quad \mathbf{v} =  |\mathbf{g}|^{-1/d} \tilde{\mathbf{v}} , \quad p =  |\mathbf{g}|^{-(d + 2)/(2 d) } \tilde{p} ,
\end{equation}
along with a rescaled time coordinate,

\begin{equation}
 \tau = \int  |\mathbf{g}|^{-1/d} \, d t.
\end{equation} 
The Lagrangian time derivative rescales as previously, with $D = |\mathbf{g}|^{-1/d} \tilde{D}$ . The fluid equations then become

\begin{align}
 \tilde{D} \tilde{\rho} &= - \tilde{\rho} \, \boldsymbol{\partial} \cdot \tilde{\mathbf{v}} , \\
 \tilde{\rho} \tilde{D} \tilde{\mathbf{v}} + \tilde{\rho} \boldsymbol{\gamma}^{-1}  \boldsymbol{\gamma}_{\tau} \tilde{\mathbf{v}} &= - \boldsymbol{\gamma}^{-1} \boldsymbol{\partial} \tilde{p} \\
\tilde{D} \tilde{p} + \gamma \tilde{p} \, \boldsymbol{\partial} \cdot \tilde{\mathbf{v}} &= \Gamma, 
\end{align}
where the effective heating/cooling rate is

\begin{equation}
 \Gamma  = -\frac{1}{2} \left( \gamma -\frac{d+2}{d} \right) \frac{\partial_{\tau} |\mathbf{g}| }{|\mathbf{g}|} \tilde{p} ,
\end{equation}
from which we see that the effective heating/cooling due to the rescaling vanishes for monatomic gases in the appropriate number of dimensions, as $\gamma = \frac{d+2}{d}$ for a d-dimensional monatomic gas.

\section{Linear perturbations in the incompressible limit} \label{linear incomp}

In this appendix we look at the linear modes of the incompressible local model equation derived in Section \ref{incomp limit}. We shall show that the anisotropy generated by the background flow induces coupling between the vortical waves and a 3rd wave responsible for enforcing incompressibility, similar to the coupling between the vortical and sound waves in the compressible case shown in \citet{Lynch23}. The linear limit of the incompressible local model ( Equations \ref{continuity old p} and \ref{momentum old p}) are

\begin{align}
  (\boldsymbol{\partial} \cdot \mathbf{v}) &= 0 , \\
  \dot{\mathbf{v}}  + \boldsymbol{\gamma}^{-1} \boldsymbol{\gamma}_{\tau} \mathbf{v} &= -  \boldsymbol{\gamma}^{-1} \boldsymbol{\partial} p , \label{linear incomp euler}
\end{align}
where, in this appendix, we have suppress the $^*$ notation on the variables. Fourier transforming these equations in space we have

\begin{align}
  \mathbf{k} \cdot \hat{\mathbf{v}} &= 0 , \\
  \hat{\mathbf{v}}_{\tau}  + \boldsymbol{\gamma}^{-1} \boldsymbol{\gamma}_{\tau} \hat{\mathbf{v}} &= -  \boldsymbol{\gamma}^{-1} i \mathbf{k} \hat{p} , 
\end{align}
By treating $p$ as a Lagrange multiplier (following the classical Lagrangian formulations for incompressible hydrodynamics \citet{Lagange88_multiplier}) the above equations can be obtained from the following Lagrangian

\begin{equation}
 L =  \frac{1}{2} \dot{\boldsymbol{\xi}}^{T} \boldsymbol{\gamma} \dot{\boldsymbol{\xi}} - i (\mathbf{k} \cdot \boldsymbol{\xi}) \hat{p}  ,
\end{equation}
where $\hat{\mathbf{v}} = \dot{\boldsymbol{\xi}}$ and a superscript $T$ denotes a transpose. The $\boldsymbol{\gamma}^{-1}$ in the pressure gradient in Equation \ref{linear incomp euler} arises from the $\boldsymbol{\gamma}$ in the kinetic energy term. Introducing the canonical momenta $\boldsymbol{\pi} = \boldsymbol{\gamma} \dot{\boldsymbol{\xi}}$, we shall solve the dynamics as a constrained Hamiltonian system \citep[e.g. see][]{Dirac50,Brown22}. The primary Hamiltonian, $H_p$ can be obtained via a Legendre transform,

\begin{equation}
 H_p = \frac{1}{2} \boldsymbol{\pi}^{T} \boldsymbol{\gamma}^{-1} \boldsymbol{\pi}  + i (\mathbf{k} \cdot \boldsymbol{\xi}) \hat{p}  .
\end{equation}
Associated with this Hamiltonian is the Poisson bracket,

\begin{equation}
 \{ f, g \} = \frac{\partial f}{\partial \boldsymbol{\xi}} \cdot \frac{\partial g}{\partial \boldsymbol{\pi}} - \frac{\partial f}{\partial \boldsymbol{\pi}} \cdot \frac{\partial g}{\partial \boldsymbol{\xi}} .
\end{equation}
We have a primary constraint

\begin{equation}
 \mathbf{k} \cdot \boldsymbol{\xi} = 0 ,
\end{equation}
requiring its time derivative to be zero we obtain the secondary constraint

\begin{equation}
 d_t ( \mathbf{k} \cdot \boldsymbol{\xi} ) = \{ \mathbf{k} \cdot \boldsymbol{\xi} , H_p\}  = \mathbf{k}^{T} \boldsymbol{\gamma}^{-1} \boldsymbol{\pi} = 0 ,
\end{equation}
where $d_t$ denotes a total time derivative. Again requiring its time derivative to be zero we obtain an  equation for the Lagrange multiplier,

\begin{align}
\begin{split}
 d_t ( \mathbf{k}^{T} \boldsymbol{\gamma}^{-1} \boldsymbol{\pi} ) &= \{ \mathbf{k}^{T} \boldsymbol{\gamma}^{-1} \boldsymbol{\pi} , H_p\}  - \mathbf{k}^{T} \boldsymbol{\gamma}^{-1} \dot{\boldsymbol{\gamma}} \boldsymbol{\gamma}^{-1} \boldsymbol{\pi} \\
&= -  i k^2 \hat{p} - \mathbf{k}^{T} \boldsymbol{\gamma}^{-1} \dot{\boldsymbol{\gamma}} \boldsymbol{\gamma}^{-1} \boldsymbol{\pi}  = 0 \quad ,
\end{split}
\end{align}
where we have introduced the wavenumber magnitude $k = \sqrt{\mathbf{k}^{T} \boldsymbol{\gamma}^{-1} \mathbf{k}}$ and obtain

\begin{equation}
   \hat{p} = \frac{i}{k^2} \mathbf{k}^{T} \boldsymbol{\gamma}^{-1} \dot{\boldsymbol{\gamma}} \boldsymbol{\gamma}^{-1} \boldsymbol{\pi} .
\end{equation}
This results in the following total Hamiltonian,

\begin{equation}
 H_T = \frac{1}{2} \boldsymbol{\pi}^{T} \boldsymbol{\gamma}^{-1} \boldsymbol{\pi} -  \frac{(\mathbf{k} \cdot \boldsymbol{\xi})}{k^2} \mathbf{k}^{T} \boldsymbol{\gamma}^{-1} \dot{\boldsymbol{\gamma}} \boldsymbol{\gamma}^{-1} \boldsymbol{\pi} \quad ,
\end{equation}
where the initial conditions are required to satisfy $(\mathbf{k} \cdot \boldsymbol{\xi})_0 = (\mathbf{k}^{T} \boldsymbol{\gamma}^{-1} \boldsymbol{\pi} )_0 = 0$. As with the compressible case considered by \citet{Lynch23}, this Hamiltonian only depends on the coordinate through $\mathbf{k} \cdot \boldsymbol{\xi}$, meaning it is possible to find new coordinates, such that we have two integrals of motion. These should correspond to the incompressible versions of the vortical-waves of \citet{Lynch23}. Additionally, when $\boldsymbol{\gamma}$ is constant the system of equations is identical to Euclidean incompressible-fluid dynamics up to a coordinate transform, we therefore wish to choose variables such that the component waves decouple. Similar to \citet{Lynch23}, to achieve this we perform a canonical transformation using the following generating function

\begin{equation}
 G = P_{\alpha} (\mathbf{k} \cdot \boldsymbol{\xi}) + P_{\beta} (k_3 \gamma_{1 0} \xi^{1} - k_1 \gamma_{3 0} \xi^{3}) + P_{\gamma} (k_1 \gamma_{2 0} \xi^{2} - k_2 \gamma_{1 0} \xi^{1}) .
\end{equation}
This results in the following relations between the old and new coordinates and momenta,

\begin{align}
 \pi_x &= P_{\alpha} k_1 + P_{\beta} k_3 \gamma_{1 0} - P_{\gamma} k_2 \gamma_{1 0}, \\
 \pi_y &= P_{\alpha} k_2 + P_{\gamma} k_1 \gamma_{2 0}, \\
 \pi_{\tilde{z}} &= P_{\alpha} k_3 - P_{\beta} k_1 \gamma_{3 0} , \\
 \alpha &= \mathbf{k} \cdot \boldsymbol{\xi} , \\
 \beta  &= k_3 \gamma_{1 0} \xi^{1} - k_1 \gamma_{3 0} \xi^{3} , \\
 \gamma &= k_1 \gamma_{2 0} \xi^{2} - k_2 \gamma_{1 0} \xi^{1} .
\end{align}
Note $\gamma$ is used here to denote one of the new coordinates, and should not be confused with the ratio of specific heats. As $G$ has no explicit dependence on time the Hamiltonian is left unchanged by this transform. To write it in the new variables we need the expression for the kinetic energy term,

\begin{align}
\begin{split}
\frac{1}{2} & \boldsymbol{\pi}^{T} \boldsymbol{\gamma}^{-1} \boldsymbol{\pi} =  k_1 k_3 \left( \frac{\gamma_{1 0}}{\gamma_{1}} - \frac{\gamma_{3 0}}{\gamma_{3}} \right) P_{\alpha} P_{\beta} \\
&+ k_1 k_2 \left( \frac{\gamma_{2 0}}{\gamma_{2}} - \frac{\gamma_{1 0}}{\gamma_{1}} \right) P_{\alpha} P_{\gamma} - k_2 k_3 \frac{\gamma_{1 0}^2}{\gamma_1} P_{\beta} P_{\gamma}  \\
&+\frac{1}{2} \left[ k^2 P_{\alpha}^2 + \left(\frac{k_3^2 \gamma_{1 0}^2}{\gamma_1} + \frac{k_1^2 \gamma_{3 0}^2}{\gamma_3} \right) P_{\beta}^2 + \left( \frac{k_2^2 \gamma_{1 0}^2}{\gamma_1} + \frac{k_1^2 \gamma_{2 0}^2}{\gamma_2}  \right) P_{\gamma}^2 \right] ,
\end{split} 
\end{align}
while $\mathbf{k}^{T} \boldsymbol{\gamma}^{-1} \dot{\boldsymbol{\gamma}} \boldsymbol{\gamma}^{-1} \boldsymbol{\pi}$, when written in terms of the new coordinate/momenta, is

\begin{align}
\begin{split}
\mathbf{k}^{T} & \boldsymbol{\gamma}^{-1} \dot{\boldsymbol{\gamma}} \boldsymbol{\gamma}^{-1}  \boldsymbol{\pi} = - \partial_{t} k^2 P_{\alpha} \\
&- k_1 k_3 \partial_t \left(\frac{\gamma_{1 0}}{\gamma_1} - \frac{\gamma_{3 0}}{\gamma_{3}}\right) P_{\beta} - k_1 k_2 \partial_t  \left(\frac{\gamma_{2 0}}{\gamma_2} - \frac{\gamma_{1 0}}{\gamma_{1}}\right) P_{\gamma} .
\end{split} 
\end{align}
This leads to a new total Hamiltonian, $K_T$, of

\begin{align}
\begin{split}
 K_T &=  \frac{1}{2} \left[ k^2 P_{\alpha}^2 + \left(\frac{k_3^2 \gamma_{1 0}^2}{\gamma_1} + \frac{k_1^2 \gamma_{3 0}^2}{\gamma_3} \right) P_{\beta}^2 + \left( \frac{k_2^2 \gamma_{1 0}^2}{\gamma_1} + \frac{k_1^2 \gamma_{2 0}^2}{\gamma_2}  \right) P_{\gamma}^2 \right] \\
 &+  k_1 k_3 \left( \frac{\gamma_{1 0}}{\gamma_{1}} - \frac{\gamma_{3 0}}{\gamma_{3}} \right) P_{\alpha} P_{\beta} + k_1 k_2 \left( \frac{\gamma_{2 0}}{\gamma_{2}} - \frac{\gamma_{1 0}}{\gamma_{1}} \right) P_{\alpha} P_{\gamma} \\
&- k_2 k_3 \frac{\gamma_{1 0}^2}{\gamma_1} P_{\beta} P_{\gamma} + \alpha \partial_{t} \ln k^2 P_{\alpha}\\
& + \alpha \left[ \frac{k_1 k_3}{k^2} \partial_t \left(\frac{\gamma_{1 0}}{\gamma_1} - \frac{\gamma_{3 0}}{\gamma_{3}}\right) P_{\beta} + \frac{k_1 k_2}{k^2} \partial_t  \left(\frac{\gamma_{2 0}}{\gamma_2} - \frac{\gamma_{1 0}}{\gamma_{1}}\right) P_{\gamma} \right] .
\end{split} 
\end{align}

We thus have the following set of evolutionary equations for the new variables

\begin{align}
 &\dot{P}_{\alpha} = - \partial_{t} \ln k^2 P_{\alpha} -  \frac{k_1 k_3}{k^2} \partial_t \left(\frac{\gamma_{1 0}}{\gamma_1} - \frac{\gamma_{3 0}}{\gamma_{3}}\right) P_{\beta} - \frac{k_1 k_2}{k^2} \partial_t  \left(\frac{\gamma_{2 0}}{\gamma_2} - \frac{\gamma_{1 0}}{\gamma_{1}}\right) P_{\gamma}   , \label{alpha wave momentum eq} \\
 &\dot{P}_{\beta} = \dot{P}_{\gamma} =  0 , \\
 &\dot{\alpha} = k^2 P_{\alpha} + k_1 k_3 \left( \frac{\gamma_{1 0}}{\gamma_{1}} - \frac{\gamma_{3 0}}{\gamma_{3}} \right)  P_{\beta} + k_1 k_2 \left( \frac{\gamma_{2 0}}{\gamma_{2}} - \frac{\gamma_{1 0}}{\gamma_{1}} \right)  P_{\gamma} + \alpha \partial_{t} \ln k^2  ,
\end{align}
where we have dropped the evolutionary equations for $\beta$ and $\gamma$ as these are shear displacements that do not affect the resulting fluid velocity or pressure. As with the fully compressible linearised equations $P_{\beta}$ and $P_{\gamma}$ are integrals of motion that are related to the conserved fluid vorticity.

To satisfy the incompressibility constraints we require $\alpha=\dot{\alpha} = 0$. From which we obtain an expression for $P_{\alpha}$ in terms of $P_{\beta}$, $P_{\gamma}$ and geometric quantities,

\begin{equation}
 P_{\alpha} = - \frac{k_1 k_3}{k^2} \left( \frac{\gamma_{1 0}}{\gamma_{1}} - \frac{\gamma_{3 0}}{\gamma_{3}} \right)  P_{\beta} - \frac{k_1 k_2}{k^2} \left( \frac{\gamma_{2 0}}{\gamma_{2}} - \frac{\gamma_{1 0}}{\gamma_{1}} \right)  P_{\gamma} .
\end{equation}
Differentiating this, noting that $P_{\beta}$ and $P_{\gamma}$ are integrals of motion, we recover Equation \ref{alpha wave momentum eq}, showing that we have now solved the linearised equations.

The fluid velocities can now be obtained by the relationship between the old an new momenta,

\begin{align}
\begin{split}
 \hat{v}^{1} &= k_3 \frac{\gamma_{1 0}}{\gamma_1} \left[ 1 - \frac{k_1^2 \gamma_{1 0}^{-1}}{k^2} \left(\frac{\gamma_{1 0}}{\gamma_{1}} - \frac{\gamma_{3 0}}{\gamma_{3}} \right)\right] P_{\beta} \\
&- k_2 \frac{\gamma_{1 0}}{\gamma_{1}} \left[ 1 + \frac{k_1^2 \gamma_{1 0}^{-1} }{k^2} \left(\frac{\gamma_{2 0}}{\gamma_2} - \frac{\gamma_{1 0}}{\gamma_{0}}\right) \right] P_{\gamma} ,
\end{split}\\
 \hat{v}^{2} &= -\frac{k_1 k_2 k_3}{k^2 \gamma_2} \left (\frac{\gamma_{1 0}}{\gamma_{1}} - \frac{\gamma_{3 0}}{\gamma_3} \right) P_{\beta} + k_1 \frac{\gamma_{2 0}}{\gamma_{2}} \left [1 - \frac{k_2^2 \gamma_{2 0}^{-1}}{k^2} \left(\frac{\gamma_{2 0}}{\gamma_{2}} - \frac{\gamma_{1 0}}{\gamma_1} \right) \right] P_{\gamma} ,\\
 \hat{v}^{3} &= -k_1 \frac{\gamma_{3 0}}{\gamma_{3}} \left[ 1 + \frac{k_3^2 \gamma_{3 0}^{-1}}{k^2} \left(\frac{\gamma_{1 0}}{\gamma_{1}} - \frac{\gamma_{3 0}}{\gamma_{3}} \right)\right] P_{\beta} - \frac{k_1 k_2 k_3}{k^2 \gamma_3} \left( \frac{\gamma_{2 0}}{\gamma_{2}} - \frac{\gamma_{1 0}}{\gamma_{1}}\right) P_{\gamma} .
\end{align}
As $P_{\beta}$ and $P_{\gamma}$ are integrals of motion, and $\gamma_{i} (0) = \gamma_{i 0}$ by definition, one can set them using the initial conditions for $\hat{\mathbf{v}}$,

\begin{align}
 \hat{v}^{1}_0 &= k_{3} P_{\beta} - k_{2} P_{\gamma}  , \\
 \hat{v}^{2}_0 &=  k_1 P_{\gamma}  , \\
 \hat{v}^{3}_0 &= - k_{1} P_{\beta}   .
\end{align}
This leads to the following solutions to the Fourier components of the linearised fluid velocities, given their initial values,

\begin{align}
 \hat{v}^{1} &= \frac{\gamma_{1 0}}{\gamma_{1}} \hat{v}_0^{1} - \frac{k_1 k_2}{k^2 \gamma_1} \left(\frac{\gamma_{2 0}}{\gamma_2} - \frac{\gamma_{1 0}}{\gamma_1} \right)\hat{v}_0^2 + \frac{k_1 k_3}{k^2 \gamma_{1}} \left(\frac{\gamma_{1 0}}{\gamma_{1}} - \frac{\gamma_{3 0}}{\gamma_{3}} \right) \hat{v}_0^3 , \label{vx lin} \\
 \hat{v}^{2} &= \frac{\gamma_{2 0}}{\gamma_{2}} \left[1 - \frac{k_2^2 \gamma_{2 0}^{-1}}{k^2} \left( \frac{\gamma_{2 0}}{\gamma_{2}} - \frac{\gamma_{1 0}}{\gamma_{1}}\right) \right] \hat{v}_0^{2} + \frac{k_2 k_3}{k^2 \gamma_2} \left(\frac{\gamma_{1 0}}{\gamma_{1}} - \frac{\gamma_{3 0}}{\gamma_{3}} \right) \hat{v}_0^{3} , \\
 \hat{v}^{3} &= - \frac{k_2 k_3}{k^2 \gamma_{3}} \left(\frac{\gamma_{2 0}}{\gamma_{\gamma_{2}}} - \frac{\gamma_{1 0}}{\gamma_{1}} \right) \hat{v}_0^2 + \frac{\gamma_{3 0}}{\gamma_3} \left[ 1 + \frac{k_3^2 \gamma_{3 0}^{-1}}{k^2} \left(\frac{\gamma_{1 0}}{\gamma_{1}} - \frac{\gamma_{3 0}}{\gamma_{3}} \right)\right] \hat{v}_0^3  . \label{vz lin}
\end{align}


\bsp	
\label{lastpage}
\end{document}